\documentclass[%
reprint,
superscriptaddress,
showpacs,%
showkeys,
amsmath,amssymb,
pre,
longbibliography, 
floatfix,
]{revtex4-1}

\usepackage[utf8]{inputenc}
\usepackage[T1]{fontenc}
\usepackage[english]{babel}
\usepackage{graphicx}
\usepackage{pstricks}
\usepackage{srcltx}

\newcommand{\diagmat}{\mathop{\text{diag}}}
\newcommand{\spanspc}{\mathop{\text{span}}}
\newcommand{\dimension}{\mathop{\text{dim}}}

\newcommand{\myexp}[1]{\mathrm{e}^{#1}}

\newcommand{\mycolon}{\,\text{:}\,}
\newcommand{\supT}{\text{T}}

\newcommand{\mat}[1]{\mbox{\boldmath{$\mathrm{#1}$}}}
\renewcommand{\vec}[1]{\mbox{\boldmath{$#1$}}}

\newcommand{\supF}{\mathcal{F}}
\newcommand{\supG}{\mathcal{G}}
\newcommand{\fpropag}{\mat{\mathcal{F}}}
\newcommand{\gpropag}{\mat{\mathcal{G}}}
\newcommand{\fsplitminus}{S^-}
\newcommand{\fsplitplus}{S^+}
\newcommand{\gsplitminus}{H^-}
\newcommand{\gsplitplus}{H^+}

\newcommand{\lsingvec}{\vec{f}^-}
\newcommand{\rsingvec}{\vec{f}^+}

\newcommand{\lsingmat}{\mat{F}^-}
\newcommand{\rsingmat}{\mat{F}^+}

\newcommand{\bkwvec}{\vec{\varphi}^-}
\newcommand{\fwdvec}{\vec{\varphi}^+}
\newcommand{\fwbkvec}{\vec{\varphi}^\pm}
\newcommand{\imgfwdvec}{\vec{\psi}^+}

\newcommand{\bkwmat}{\mat{\Phi}^-} 
\newcommand{\fwdmat}{\mat{\Phi}^+}
\newcommand{\imgfwdmat}{\mat{\Psi}^+}

\newcommand{\clvecfwd}{\vec{\gamma}}
\newcommand{\clmatfwd}{\mat{\Gamma}}
\newcommand{\clvecadj}{\vec{\theta}}
\newcommand{\clmatadj}{\mat{\Theta}} 
\newcommand{\identmat}{\mat{I}}
\newcommand{\singloclyap}{\tilde{\mu}}
\newcommand{\loclyap}{\tilde{\lambda}}
\newcommand{\stpcomp}{K_{\text{AB}}}
\newcommand{\stptran}{K_{\text{BC}}}
\newcommand{\stptrancoef}{k_{\text{BC}}}
\newcommand{\fundmat}{\mat{M}}
\newcommand{\signat}{\mat{D}}
\newcommand{\matcfwd}{\mat{C}^\supF}
\newcommand{\matcfwdelem}{c^\supF}
\newcommand{\matcadj}{\mat{C}^\supG}

\begin{document}

\title{Theory and computation of covariant Lyapunov
  vectors}

\author{Pavel V. Kuptsov}\email[Corresponding author. Electronic
address:]{p.kuptsov@rambler.ru}%
\affiliation{Department of Instrumentation Engineering,
  Saratov State Technical University, Politekhnicheskaya 77, Saratov
  410054, Russia}%

\author{Ulrich Parlitz}%
\affiliation{Biomedical Physics Group, Max Planck Institute for
  Dynamics and Self-Organization, Am Fa\ss berg 17, 37077 G\"{o}ttingen,
  Germany}%
\affiliation{Institute for Nonlinear Dynamics,
  Georg--August--Universit\"at G\"ottingen, Am Fassberg 17, 37077
  G\"ottingen, Germany}



\date{\today}

\begin{abstract}
  Lyapunov exponents are well-known characteristic numbers that
  describe growth rates of perturbations applied to a trajectory of a
  dynamical system in different state space directions. Covariant (or
  characteristic) Lyapunov vectors indicate these directions. Though
  the concept of these vectors has been known for a long time, they became
  practically computable only recently due to algorithms suggested by
  Ginelli et al. [Phys. Rev. Lett. 99, 2007, 130601] and by Wolfe and
  Samelson [Tellus 59A, 2007, 355]. In view of the great interest in
  covariant Lyapunov vectors and their wide range of potential
  applications, in this article we summarize the available information
  related to Lyapunov vectors and provide a detailed explanation of
  both the theoretical basics and numerical algorithms. We introduce
  the notion of adjoint covariant Lyapunov vectors. The angles between
  these vectors and the original covariant vectors are
  norm-independent and can be considered as characteristic
  numbers. Moreover, we present and study in detail an improved
  approach for computing covariant Lyapunov vectors. Also we describe,
  how one can test for hyperbolicity of chaotic dynamics without
  explicitly computing covariant vectors.
\end{abstract}

\maketitle

\section*{Introduction}

High-dimensional nonlinear systems like coupled oscillators, dynamical
networks, or extended excitable media often exhibit very complex
dynamics that is difficult to analyze and to characterize. From a
practical point of view there only a few concepts have been developed
for studying low-dimensional systems that can efficiently be applied
to high-dimensional attractors, too. An important example are Lyapunov
exponents that describe growth rates of perturbations applied to a
trajectory in different state space directions. These exponents are a
central point in the investigation of chaotic dynamical systems.  They
are related to a number of different physical properties such as
sensitivity to initial conditions or local entropy production and can
be used to estimate the (Kaplan--Yorke) dimension of (even very
high-dimensional) attractors~\cite{EckRuell85}.

Mathematically, Lyapunov exponents are defined in tangent space. This
space is spanned by all possible infinitesimal perturbations that can
be applied to a state of the system. The dimension of the tangent
space is equal to the dimension of the original phase space. In
general, the tangent space is an inner product space, but often the
tangent space is defined as an Euclidean space where the inner product
is just the ordinary scalar (dot) product.  The dynamics in this space
is generated by linear operators, that determine the evolution of
perturbation vectors from one point on the trajectory to
another. These operators are called tangent linear propagators or
resolvents. The tangent space is a very important subject of study. On
the one hand, the tangent space dynamics is closely related to
the dynamics of the original system. One can obtain key
characteristics of the original system observing the associated
tangent space dynamics. On the other hand, the tangent space is
linear and the dynamics in this space is determined by the action of
linear operators.  This means that analysis methods as well as results
are universal for a wide class of systems.

Besides the growth rates of perturbations the directions of this
growth are also important. There are various concepts identifying
these directions including bred vectors~\cite{TothKaln93,TothKaln97},
which are finite-amplitude perturbations initialized and periodically
rescaled within the original phase space, singular or optimal
vectors~\cite{Buizza93,Buizza95}, which are the singular vectors of a
finite-time propagator, or finite-time normal modes~\cite{Fred97},
defined as eigenvectors of the propagator.

Orthogonal sets of singular vectors related to the propagators
operating on infinite time intervals were referred to by Legras and
Vautard as forward and backward Lyapunov vectors~\cite{AGuide}. These
vectors can be computed in parallel with the Lyapunov
exponents~\cite{AGuide, ErshovPotapov98}, and, thus, are closely
related to them. Unlike the exponents, the forward and backward
Lyapunov vectors depend on time, i.e., they are different for different
trajectory points. Analyzing the orientation of these vectors, one can
expect to recover the local structure of an attractor. But
unfortunately, the forward and backward Lyapunov vectors provide only
limited information. They always remain orthogonal and thus cannot
indicate directions of stable and unstable manifolds as well as their
tangencies. These vectors are not invariant under time reversal and
are not covariant with the dynamics. The latter means that forward (or
backward) vectors at a given point are not mapped by tangent
propagators to the forward (backward) vectors at the image point.
Another drawback of these vectors is their norm-dependence, i.e., they
depend on the definition of the inner products and norms in the
tangent space~\cite{AGuide}.

The concept of norm-independent Lyapunov vectors ha been known for a
long time~\cite{EckRuell85,VastMoser91,AGuide,TrevPan98}. However,
only recently two efficient algorithms for computing these vectors
were suggested almost simultaneously by Wolfe and
Samelson~\cite{WolfCLV} and by Ginelli et al.~\cite{GinCLV}. After
Ginelli et al. we call these vectors covariant Lyapunov vectors. Note
that these vectors are also referred to as characteristic Lyapunov
vectors~\cite{AGuide,WolfCLV}. These vectors are not orthogonal, they
are invariant under time reversal and covariant with the dynamics in
the sense that they may, in principle, be computed once and then
determined for all times using the tangent propagator. (Note that this
is the case only for exact covariant vectors, while those computed
numerically do not demonstrate perfect covariance due to the
accumulation of numerical errors.) The covariant Lyapunov vectors can
be considered as a generalization of the notion of ``normal modes.''
They are reduced to Floquet vectors if the flow is time periodic and
to stationary normal modes if the flow is stationary~\cite{WolfCLV}.

In view of potential wide applications to the analysis of complex,
high-dimensional dynamics, the covariant Lyapunov vectors receive a
lot of interest of researchers~\cite{SzendroPazo07, HyperSpace08,
  PazoSzendro08, EffDim, ModesSplit10, YangRad2010, PazoLopez10}. For
these extensive studies to be productive, it is important to analyze
the Lyapunov vectors systematically. In this paper we summarize
features of forward, backward and covariant Lyapunov vectors and
provide a detailed explanation of both the theoretical basics and
numerical algorithms. We present and study in detail an efficient
method for computing covariant Lyapunov vectors, which can be
considered as a modification of the method by Wolfe and Samelson.
Moreover, our general approach reveals the existence of adjoint
covariant Lyapunov vectors. This is \emph{not an independent type of
  characteristic vectors}, because given the covariant vectors, one
can always compute the adjoint ones. However, the angles between
corresponding covariant and adjoint covariant vectors provide a
compact representation of the information contained in the covariant
vectors and can be used as characteristic numbers. In particular, the
presence of homoclinic tangencies is indicated by orthogonality of
corresponding original and adjoint covariant vectors. Since the
covariant as well as the adjoint covariant vectors are
norm-independent their angles also are invariant with respect to the
norm.

The structure of the article is as follows. In
Sec.~\ref{sec:lyap-expon-lyap} we present the theory of Lyapunov
exponents and forward and backward Lyapunov vectors, and in
Sec.~\ref{sec:num_comp_lyap} we describe numerical methods for
computing them. Section~\ref{sec:covar-lyap-vect} presents the
theoretical aspects of covariant Lyapunov vectors, and in
Sec.~\ref{sec:pract-meth-comp} we describe different methods of
computing covariant vectors. Finally, in Sec.~\ref{sec:examp} a simple
illustrative example is presented. In Sec.~\ref{sec:conclusion} we
summarize the results presented.

\section{Lyapunov exponents, forward and backward Lyapunov
  vectors}\label{sec:lyap-expon-lyap}

\subsection{Basic definitions}\label{sec:bas-defin}
Consider a system whose dynamics can be described by an ordinary
differential equation
\begin{equation}\label{eq:basic_ode}
  \dot{\vec{u}}=\vec{g}(\vec{u},t),
\end{equation}
where $\vec{u}\equiv\vec{u}(t)\in\mathbb{R}^m$ is an $m$-dimensional
state vector that changes in time $t$, and
$\vec{g}(\vec{u},t)\in\mathbb{R}^m$ is a nonlinear vector function.
We are primarily interested in high-dimensional systems, so $m$ is
assumed to be large. Equation~\eqref{eq:basic_ode} can model a system
with many interacting point-wise dynamical elements, or it can be a
finite step size approximation of a spatially extended system that
appears after discretization of spatial derivatives. Infinitesimal
perturbations to a trajectory of this system are described by the
following equation:
\begin{equation}\label{eq:basic_ode_lin}
  \dot{\vec{v}}=\mat{J}(\vec{u},t) \vec{v},
\end{equation}
where $\mat{J}(\vec{u},t)\in\mathbb{R}^{m\times m}$ is the Jacobian
matrix composed of derivatives of the vector function
$\vec{g}(\vec{u},t)$ with respect to components of the vector
$\vec{u}$. The fundamental matrix $\fundmat\in\mathbb{R}^{m\times m}$
for Eq.~\eqref{eq:basic_ode_lin} can be found as a solution of the
matrix equation
\begin{equation}\label{eq:fundam_syst}
  \dot{\fundmat}=\mat{J}(\vec{u},t) \fundmat,
\end{equation}
where any non-singular matrix can be used as an initial condition.

The tangent linear propagator or resolvent is defined as
\begin{equation}\label{eq:propagator}
  \fpropag(t_1,t_2)=\fundmat(t_2)\fundmat(t_1)^{-1},
\end{equation}
and can be represented by a  non-singular $m\times m$ matrix.
The propagator evolves solutions of Eq.~\eqref{eq:basic_ode_lin} from
time $t_1$ to time $t_2$:
\begin{equation}\label{eq:propagator_action}
  \vec{v}(t_2)=\fpropag(t_1,t_2)\vec{v}(t_1),
\end{equation}
where $\vec{v}(t_1)$ and $\vec{v}(t_2)$ are tangent vectors at times
$t_1$ and $t_2$, respectively, computed along the same trajectory of
the base system~\eqref{eq:basic_ode}. According to
Eq.~\eqref{eq:propagator}, the propagator is always non-singular and
$\fpropag(t_1,t_2)=\fpropag(t_2,t_1)^{-1}$. Furthermore we define the
adjoint tangent propagator:
\begin{equation}\label{eq:gpropag_defin}
  \gpropag(t_1,t_2)=\fpropag(t_1,t_2)^{-\supT},
\end{equation}
where ``$-\supT$'' denotes matrix inversion and transposition. In
general, a non-Euclidean norm can be defined in the tangent space, so
that instead of the transposition a generalized adjoint with respect
to the chosen norm has to be used. In this paper we do not consider
such cases.

As follows from Eq.~\eqref{eq:propagator_action}, the growth of the
Euclidean norm of tangent vectors in forward-time dynamics is
determined by the matrix $\fpropag(t_1,t_2)^\supT
\fpropag(t_1,t_2)$. We denote its eigenvectors and eigenvalues as
$\rsingvec_i(t_1,t_2)$ and $\sigma_i(t_1,t_2)^2$, respectively, where
$\sigma_1(t_1,t_2) \geq \sigma_2(t_1,t_2) \geq \cdots \geq
\sigma_m(t_1,t_2)\geq 0$. The eigenvectors are termed optimal vectors
because the maximal growth ratio is equal to $\sigma_1(t_1,t_2)$ and
is achieved if the initial vector $\vec{v}(t_1)$ coincides with
$\rsingvec_1(t_1,t_2)$. The same role for the backward-time dynamics
plays the matrix $\fpropag(t_1,t_2)^{-\supT} \fpropag(t_1,t_2)^{-1}$
with the reciprocal eigenvalues and the eigenvectors
$\lsingvec_i(t_1,t_2)$.

The eigenvectors and eigenvalues can be found via singular value
decompositions (SVD) \cite{GolubLoan} of the propagator matrix and its
inverse, thus:
\begin{align}
  \label{eq:fsvd_fwd}
  \fpropag(t_1,t_2)\rsingvec_i(t_1,t_2)&=
  \lsingvec_i(t_1,t_2)\sigma_i(t_1,t_2),\\
  \label{eq:fsvd_bkw}
  \fpropag(t_1,t_2)^{-1}\lsingvec_i(t_1,t_2)&=
  \rsingvec_i(t_1,t_2)\sigma_i(t_1,t_2)^{-1}.
\end{align}
Here $\sigma_i(t_1,t_2)$ are called singular values, and
$\rsingvec_i(t_1,t_2)$ and $\lsingvec_i(t_1,t_2)$ are right and left
singular vectors of $\fpropag(t_1,t_2)$, respectively. The singular
vectors are orthonormal. They are norm-dependent, i.e., they have
different orientations with respect to different
norms~\cite{AGuide,WolfCLV}. Taking into account
Eqs.~\eqref{eq:gpropag_defin}, \eqref{eq:fsvd_fwd}, and
\eqref{eq:fsvd_bkw}, one can write the SVD for the adjoint propagator
$\gpropag(t_1,t_2)$ and its inverse as
\begin{align}
  \label{eq:gsvd_fwd}
  \gpropag(t_1,t_2)\rsingvec_i(t_1,t_2)&=
  \lsingvec_i(t_1,t_2)\sigma_i(t_1,t_2)^{-1},\\
  \label{eq:gsvd_bkw}
  \gpropag(t_1,t_2)^{-1}\lsingvec_i(t_1,t_2)&= 
  \rsingvec_i(t_1,t_2)\sigma_i(t_1,t_2).
\end{align}
Comparing Eq.~\eqref{eq:fsvd_fwd} with \eqref{eq:gsvd_fwd} and
Eq.~\eqref{eq:fsvd_bkw} with \eqref{eq:gsvd_bkw} we see that the
propagators $\fpropag(t_1,t_2)$ and $\gpropag(t_1,t_2)$ have identical
singular vectors and reciprocal singular values.

If all $\sigma_i(t_1,t_2)$ are distinct, the singular vectors are
unique up to a simultaneous change of signs of elements of
$\rsingvec_i(t_1,t_2)$ and $\lsingvec_i(t_1,t_2)$. In the presence of
degeneracy, we still can find a set of orthonormal right singular
vectors that are mapped according to Eq.~\eqref{eq:fsvd_fwd} onto a
set of orthonormal left singular vectors, but these sets are not
unique and can be selected arbitrarily.

Strictly speaking, propagators and singular vectors as well as the
 Lyapunov vectors considered below can depend on time both explicitly,
and implicitly via state vectors $\vec{u}(t)$.  To avoid 
complicated notation, we shall use  a compact form, like
$\fpropag(t_1,t_2)$.

\subsection{Properties of propagators. Transformation of volumes built
  on singular vectors}

\begin{figure*}
  \begin{center}
    a)\ifx\JPicScale\undefined\def\JPicScale{1}\fi
\psset{unit=\JPicScale mm}
\psset{linewidth=0.3,dotsep=1,hatchwidth=0.3,hatchsep=1.5,shadowsize=1,dimen=middle}
\psset{dotsize=0.7 2.5,dotscale=1 1,fillcolor=black}
\psset{arrowsize=1 2,arrowlength=1,arrowinset=0.25,tbarsize=0.7 5,bracketlength=0.15,rbracketlength=0.15}
\begin{pspicture}(0,0)(121.72,58.91)
\pspolygon[linewidth=0.15,fillstyle=vlines,hatchwidth=0.25,hatchsep=2](68.24,28.27)(77.82,10.71)(121.72,34.65)(112.14,52.21)
\pspolygon[linewidth=0.15,fillstyle=vlines,hatchwidth=0.25,hatchsep=2](21.66,50.1)(48.49,36.69)(35.08,9.85)(8.25,23.26)
\psline[linewidth=0.45,arrowscale=1.6 1.8]{->}(8.24,23.27)(21.66,50.1)
\psline[linewidth=0.45,arrowscale=1.6 1.8]{->}(8.24,23.27)(35.08,9.85)
\psline[linewidth=0.45,arrowscale=1.6 1.8]{->}(68.24,28.27)(112.14,52.21)
\psline[linewidth=0.45,arrowscale=1.6 1.8]{->}(68.24,28.27)(77.82,10.71)
\rput{0}(59.77,25.36){\psellipticarc[linewidth=0.7,arrowscale=1.45 2.5]{<-}(0,0)(30.54,30.54){33.44}{137.1}}
\rput[B](66.66,58.91){$\fpropag(t_1,t_2)$}
\rput[br](21.92,52.03){$\rsingvec_1$}
\rput[br](113.33,53.97){$\sigma_1\lsingvec_1$}
\rput[tr](78.04,8.3){$\sigma_2\lsingvec_2$}
\rput[tr](35.68,7.74){$\rsingvec_2$}
\end{pspicture}\\[0.5cm]
    b)\ifx\JPicScale\undefined\def\JPicScale{1}\fi
\psset{unit=\JPicScale mm}
\psset{linewidth=0.3,dotsep=1,hatchwidth=0.3,hatchsep=1.5,shadowsize=1,dimen=middle}
\psset{dotsize=0.7 2.5,dotscale=1 1,fillcolor=black}
\psset{arrowsize=1 2,arrowlength=1,arrowinset=0.25,tbarsize=0.7 5,bracketlength=0.15,rbracketlength=0.15}
\begin{pspicture}(0,0)(116.45,64.36)
\psline[linewidth=0.45,arrowscale=1.6 1.8]{->}(75.77,33.54)(90.09,7.16)
\rput{0}(57.49,26.57){\psellipticarc[linewidth=0.7,arrowscale=1.45 2.5]{->}(0,0)(34.33,34.33){27.94}{145.03}}
\rput[B](60.37,64.36){$\fpropag(t_1,t_2)^{-1}$}
\rput[br](20.24,52.34){$\sigma_1^{-1}\rsingvec_1$}
\rput[bl](100.9,49.49){$\lsingvec_1$}
\rput[tr](87.78,6.45){$\lsingvec_2$}
\pspolygon[linewidth=0.15,fillstyle=vlines,hatchwidth=0.25,hatchsep=2](102.14,47.84)(116.45,21.48)(90.09,7.16)(75.78,33.52)
\rput[tr](49.49,10.19){$\sigma_2^{-1}\rsingvec_2$}
\psline[linewidth=0.45,arrowscale=1.6 1.8]{->}(75.77,33.54)(102.14,47.84)
\pspolygon[linewidth=0.15,fillstyle=vlines,hatchwidth=0.25,hatchsep=2](6.47,32.98)(46.92,12.59)(55.94,30.47)(15.49,50.86)
\psline[linewidth=0.45,arrowscale=1.6 1.8]{->}(6.56,33.22)(15.45,50.63)
\psline[linewidth=0.45,arrowscale=1.6 1.8]{->}(6.56,33.22)(46.79,12.72)
\end{pspicture}
  \end{center}
  \caption{Transformation of a volume. (a) Forward step by the
    propagator $\fpropag(t_1,t_2)$, (b) backward step via
    $\fpropag(t_1,t_2)^{-1}$.}
  \label{fig:phase_vol_transf}
\end{figure*}
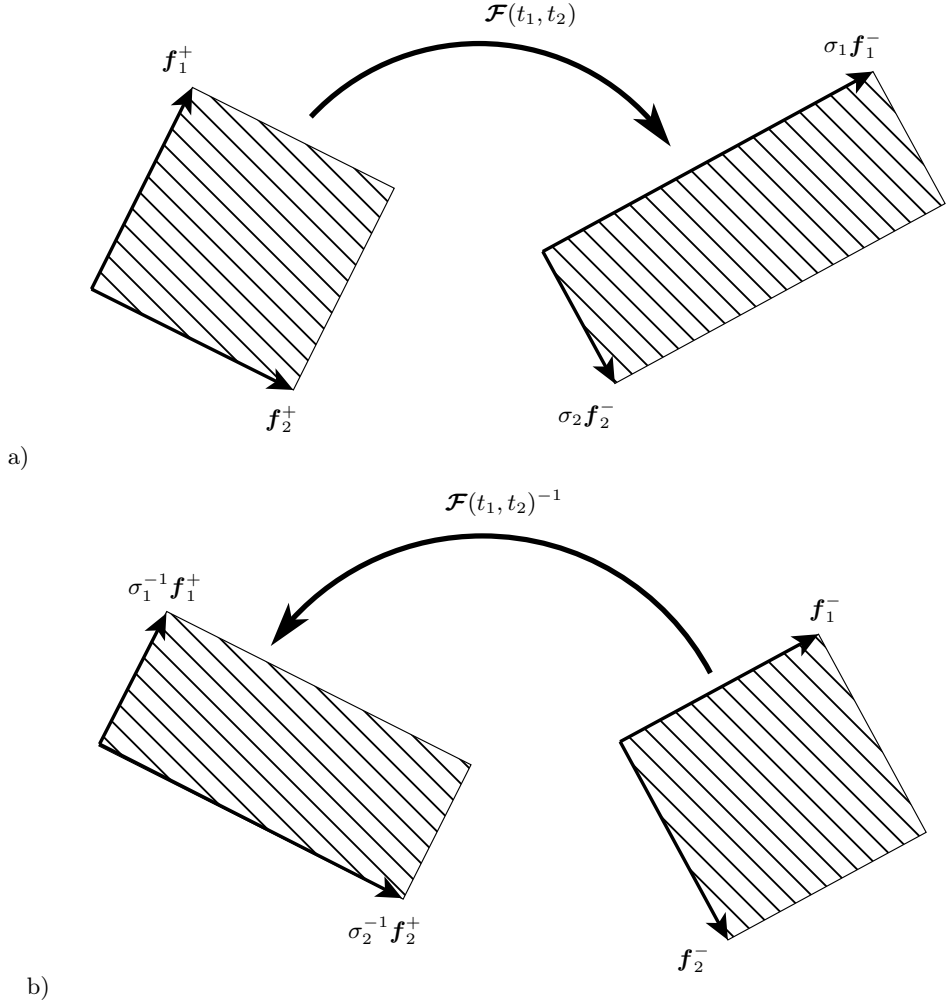

Let us discuss how $\fpropag(t_1,t_2)$ transforms volumes of different
dimensions: segments, squares, cubes and so on. Being at a trajectory
point at $t_1$ we construct a $k$-dimensional unit volume using the
first $k$ right singular vectors $\rsingvec_i(t_1,t_2)$. According to
Eq.~\eqref{eq:fsvd_fwd} $\fpropag(t_1,t_2)$ transforms these vectors
into the left singular vectors $\lsingvec_i(t_1,t_2)$ associated with
the trajectory point at $t_2$ that are stretched/contracted by factors
$\sigma_i(t_1,t_2)$, see Fig.~\ref{fig:phase_vol_transf}(a). The
volume at $t_2$ is equal to the product of the first $k$ singular
values. Alternatively, we can consider a $k$-dimensional ball of unit
radius at $t_1$. At $t_2$ this ball is transformed into an ellipsoid
with axes along the vectors $\lsingvec_i$ and lengths $\sigma_i$. One
can describe this transformation of volumes by
\begin{equation}\label{eq:volume_transf}
  V_k(t_2)=V_k(t_1)\exp\left( (t_2-t_1)
    \sum_{i=1}^k \singloclyap_i(t_1,t_2) \right),
\end{equation}
where $V_k(t)$ is the $k$-dimensional volume, and
$\singloclyap_i(t_1,t_2)=\ln \sigma_i(t_1,t_2) / (t_2-t_1)$ are
stretch ratios that can be considered as local Lyapunov
exponents. (Note that there are alternative definitions of local
Lyapunov exponents that shall be considered below.)

The backward transformation with $\fpropag(t_1,t_2)^{-1}$ is
symmetric. At $t=t_2$ we construct a unit volume using the first $k$
left singular vectors $\lsingvec_i(t_1,t_2)$.  According to
Eq.~\eqref{eq:fsvd_bkw}, the right singular vectors span this volume
at $t=t_1$, and the edges of this volume are stretched/contracted by
factors $\sigma^{-1}_i$, see Fig.~\ref{fig:phase_vol_transf}(b). In a
similar manner we can consider a unit ball at $t_2$ that is
transformed into an ellipsoid at $t_1$. Therefore, the volumes are again
transformed in accordance with Eq.~\eqref{eq:volume_transf}.

This discussion is also valid for the adjoint propagator
$\gpropag(t_1,t_2)$. But because the singular values are now
reciprocal, the volumes are transformed as
\begin{equation}\label{eq:volume_transf_adj}
  V_k(t_2)=V_k(t_1)\exp\left(-(t_2-t_1)
    \sum_{i=1}^k \singloclyap_i(t_1,t_2) \right),
\end{equation}

\subsection{Far-past and far-future operators. Forward and backward
  Lyapunov vectors}

For infinitely large time intervals we can expect to obtain limits
for the stretch ratios and singular vectors. The Oseledec
multiplicative ergodic theorem~\cite{Oseled} and its corollaries state
that the limit indeed exists for $t_2\to\infty$, and also a  limit can
be reached for $t_1\to -\infty$. When $t_2\to\infty$, the
\emph{far-future operator} is defined as
\begin{align}
  \label{eq:far_fwd_oper}
  \mat{W}^+(t)&=\lim_{t_2\to\infty} \left[\fpropag(t,t_2)^\supT
    \fpropag(t,t_2)\right]^{1/(2(t_2-t))}\\
  {}&= \lim_{t_2\to\infty} \left[
    \rsingmat(t,t_2) \mat{\Sigma}(t,t_2)^{1/(t_2-t)}
    \rsingmat(t,t_2)^\supT \right],\nonumber
\end{align}
where
$\rsingmat(t,t_2)=[\rsingvec_1(t,t_2),\ldots,\rsingvec_m(t,t_2)]$ and
$\mat{\Sigma}(t,t_2)=\diagmat[\sigma_1(t,t_2),\ldots,
\sigma_m(t,t_2)]$ are matrices of singular vectors and values,
respectively. The eigenvectors of the far-future operator are the
limits of vectors $\rsingvec_i(t,t_2)$. We denote them as
$\fwdvec_i(t)$ and refer to them as \emph{forward Lyapunov
  vectors}. They are orthonormal and depend on $t$~\cite{AGuide}. The
convergence of the singular vectors to the Lyapunov vectors is
considered in Ref.~\cite{ReyErr99}.  Logarithms of eigenvalues of
$\mat{W}^+(t)$, $\lambda_1\geq\lambda_2\geq\cdots\geq\lambda_m$, are
called Lyapunov exponents. Regardless of time dependence of
$\mat{W}^+(t)$, they do not depend on time.

The \emph{far-past operator} is defined as
\begin{align}\label{eq:far_pst_oper}
  \mat{W}^-(t)&=\lim_{t_1\to -\infty}\left[\fpropag(t_1,t)^{-\supT}
    \fpropag(t_1,t)^{-1}\right]^{1/(2(t-t_1))}\\
  {}&=
  \lim_{t_1\to-\infty} \left[
    \lsingmat(t_1,t)
    \mat{\Sigma}(t_1,t)^{-1/(t-t_1)}
    \lsingmat(t_1,t)^\supT
  \right],\nonumber
\end{align}
where
$\lsingmat(t_1,t)=[\lsingvec_1(t_1,t),\ldots,\lsingvec_m(t_1,t)]$.
The eigenvectors of this matrix are the limits of the left singular
vectors $\lsingvec_i(t_1,t)$ for $t_1 \to -\infty$. They are called
\emph{backward Lyapunov vectors}. These vectors are also referred to
as Gram--Schmidt vectors, because they can be computed in the course of
a procedure, which includes Gram--Schmidt orthogonalizations;
see. Sec.~\ref{sec:num_comp_lyap}. We denote them by
$\bkwvec_i(t)$. Similar to the forward vectors, the backward Lyapunov
vectors are orthonormal, and depend on $t$~\cite{AGuide}. As well as
singular vectors, forward and backward Lyapunov vectors are
norm-dependent~\cite{AGuide,WolfCLV}.  The logarithms of the
eigenvalues of $\mat{W}^-(t)$ are equal to the Lyapunov exponents with
opposite signs.

In analogy with the finite-time case, the $k$-dimensional volumes can
be built on the forward Lyapunov vectors $\fwdvec_i(t)$. Modifying
Eq.~\eqref{eq:volume_transf} we find that average growth rates of
these volumes are the sums of Lyapunov exponents,
\begin{equation}\label{eq:volume_transf_lim}
  \sum_{i=1}^k \lambda_i
  =\lim_{t_2\to\infty}\left(
    \frac{1}{t_2-t_1}\ln
    \frac{V_k(t_2)}{V_k(t_1)}
  \right).
\end{equation}
As we shall see below, this formula is valid for almost any
$k$-dimensional volume in the tangent space, not necessarily related to
the forward Lyapunov vectors.

The Lyapunov exponents may not be all distinct. To take possible
degeneracy into account we introduce an additional notation. Let $s$
be a number of \emph{distinct} Lyapunov exponents ($1\leq s\leq m$),
and let $\lambda^{(i)}$ ($i=1,2,\ldots,s$) denote the $i$th distinct
Lyapunov exponent with the multiplicity $\nu^{(i)}$. So, we have
$\lambda^{(1)}>\lambda^{(2)}>\cdots>\lambda^{(s)}$, and
$\sum_{i=1}^{s}\nu^{(i)}=m$. In what follows, to address the whole
spectrum of Lyapunov exponents as well as related vectors, we shall
employ lower indices while paying special attention to the
multiplicity, we shall use upper indices. The notation
$\fwbkvec_{\lambda^{(i)}}$ will stand for a set of vectors, related to
the $i$th distinct Lyapunov exponent, and
$\fwbkvec_{\lambda^{(i)},j}$, where $j=1,2,\ldots,\nu^{(i)}$, will
denote the $j$th vector related to $\lambda^{(i)}$.

In presence of the degeneracy forward and backward Lyapunov vectors
are not unique. But as we already mentioned for singular vectors, this
is not an obstacle, because it is always sufficient to choose any
orthonormal set of these vectors.

The adjoint propagator $\gpropag$ can also be used to define far-past
and far-future operators and forward and backward vectors,
respectively. The Lyapunov exponents are the logarithms of the
eigenvalues of the far-past operator, while the far-future operator is
associated with the Lyapunov exponents with inverted signs.

\subsection{Oseledec subspaces. Asymptotic behavior of arbitrary
  vectors and volumes}

Let us now discuss what happens with arbitrary vectors. The framework
that helps to understand it is provided by the following set of
subspaces:
\begin{gather}
  \fsplitplus_j(t)=\spanspc 
  \left\{
    \fwdvec_{\lambda^{(i)}}(t)\big|i=j,j+1,\ldots,s
  \right\}, \fsplitplus_{s+1}(t)=\emptyset,\nonumber\\
  \fsplitplus_s(t)\subset\fsplitplus_{s-1}(t)\subset\cdots
  \subset\fsplitplus_1(t)=\mathbb{R}^m.
  \label{eq:fsplit_plus}
\end{gather}
In other words, $\fsplitplus_j(t)$ is spanned by forward Lyapunov
vectors $\fwdvec_{\lambda^{(i)}}(t)$ ($i\geq j$) related to the
distinct Lyapunov exponents starting from the $j$th one. Dimensions
of these subspaces are
$\dimension\fsplitplus_j(t)=\sum_{i=j}^s\nu^{(i)}$, where $\nu^{(i)}$
is the multiplicity of $\lambda^{(i)}$. Analogous subspaces spanned by
the backward Lyapunov vectors $\bkwvec_{\lambda^{(i)}}(t)$ are defined
by
\begin{gather}
  \fsplitminus_j(t)=\spanspc
  \left\{
    \bkwvec_{\lambda^{(i)}}(t)\big|i=1,2,\ldots,j
  \right\},\fsplitminus_0(t)=\emptyset,\nonumber\\
  \fsplitminus_1(t)\subset\fsplitminus_2(t)\subset\cdots
  \subset\fsplitminus_s(t)=\mathbb{R}^m,
  \label{eq:fsplit_minus}
\end{gather}
and their dimensions are
$\dimension\fsplitminus_j(t)=\sum_{i=1}^j\nu^{(i)}$.  These sets of
subspaces are referred to as Oseledec
splitting~\cite{Oseled,Benettin,AGuide}.

Recall that the propagator $\fpropag(t_1,t_2)$ maps each right
singular vector onto the corresponding left singular vector and
stretching rates are determined by singular values, see
Eq.~\eqref{eq:fsvd_fwd}. When $(t_2-t_1)\to\infty$, the right and left
singular vectors converge to forward and backward Lyapunov vectors,
respectively, and the stretching rates converge to the Lyapunov
exponents. Hence, the Oseledec subspace $\fsplitplus_j(t)$ consists
of vectors that asymptotically grow or decay with rate $\lambda\leq
\lambda^{(j)}$. In turn, the vectors from Oseledec subspace
$\fsplitminus_j(t)$ grow or decay with exponential rates $\lambda\geq
\lambda^{(j)}$ \emph{backward in time}.

Consider a vector $\vec{v}^{(j)}(t)\in\fsplitplus_j(t)\setminus
\fsplitplus_{j+1}(t)$. This vector is orthogonal to each
$\fwdvec_{\lambda^{(i)}}(t)$, where $i<j$, and obligatory has a
nonzero projection onto at least one of the vectors
$\fwdvec_{\lambda^{(j)}}(t)$, related to the $j$th distinct Lyapunov
exponent. It means that being iterated for infinitely long time with
the propagator $\fpropag$, the vector $\vec{v}^{(j)}(t)$ exponentially
grows with the average rate
$\lambda^{(j)}$~\cite{Oseled,Benettin,Shimada79},
\begin{multline}
  \label{eq:asympt_exp_ffwd}
  \vec{v}^{(j)}(t_1)\in
  \fsplitplus_j(t_1)\setminus\fsplitplus_{j+1}(t_1) %
  \Rightarrow \\
  \|\fpropag(t_1,t_1+t)\vec{v}^{(j)}(t_1)\|\sim \myexp{\lambda^{(j)} t}.
\end{multline}
The vectors $\vec{v}^{(j)}(t)\in
\fsplitminus_j(t)\setminus\fsplitminus_{j-1}(t)$ behave analogously in
backward time:
\begin{multline}
  \label{eq:asympt_exp_fbkw}
  \vec{v}^{(j)}(t_1)\in
  \fsplitminus_j(t_1)\setminus\fsplitminus_{j-1}(t_1) %
  \Rightarrow \\
  \|\fpropag(t_1-t,t_1)^{-1}\vec{v}^{(j)}(t_1)\|\sim
  \myexp{-\lambda^{(j)}t}.
\end{multline}
Vectors $\vec{v}^{(1)}(t)\in\fsplitplus_1(t)\setminus
\fsplitplus_2(t)$ fill almost the whole tangent space, because the
excluded subspace $\fsplitplus_2(t)$ has a measure zero in
$\mathbb{R}^m$. It means that under the action of $\fpropag$ almost
any vector, i.e., 1-dimensional volume, asymptotically grows or decays
with the exponent $\lambda^{(1)}$, and its image tends to the subspace
$\spanspc\{\bkwvec_{\lambda^{(1)}}(t)\} =\fsplitminus_1(t)$.  Consider
now a square, i.e., a 2-dimensional volume. First we assume that
$\lambda^{(1)}$ is not degenerate so that $\nu^{(1)}=1$. Almost any
such square has a 1-dimensional intersection with the subspace
$\fsplitplus_2(t)\setminus \fsplitplus_{3}(t)$ of vectors
$\vec{v}^{(2)}(t)$ that are dominated by the
$\lambda^{(2)}$~\cite{Benettin,Shimada79,AGuide,ParkerChua}.  (Here
``almost'' means that there is a measure zero set of squares fully
belonging to subspaces with $j>1$.) Thus, the area of the square
asymptotically grows or decays with the exponent
$\lambda^{(1)}+\lambda^{(2)}$. All segments within this square except
a single one approach the subspace $\spanspc\{
\bkwvec_{\lambda^{(1)}}(t)\}$, while that one goes into
$\spanspc\{\bkwvec_{\lambda^{(2)}}(t)\}$. As a result, this square
tends into the subspace $\fsplitminus_2$.  When $\nu^{(1)}=2$, the
area of the square grows/decays with
$2\lambda^{(1)}=\lambda_1+\lambda_2$ and the whole square is embedded
into $\fsplitminus_1$. But when we take a cube, its volume grows or
decays with
$2\lambda^{(1)}+\lambda^{(2)}=\lambda_1+\lambda_2+\lambda_3$ and its
image goes into $\fsplitminus_2$. In general this can be formulated as
follows. Under the action of $\fpropag$ almost any $k$-dimensional
volume asymptotically grows or decays with average exponential rate
$\sum_{i=1}^k\lambda_i$ and tends to settle down inside the subspace
$\fsplitminus_i$, where $i$ is defined from the inequalities
$\dimension\fsplitminus_{i-1}<k \leq \dimension\fsplitminus_{i}$.  In
the same way considering vectors $\vec{v}^{(j)}(t)
\in\fsplitminus_j(t)\setminus \fsplitminus_{j-1}(t)$ we see that
almost any $k$-dimensional volume being iterated in backward time with
the propagator $\fpropag(t_1,t_2)^{-1}$ grows or decays with the
exponential rate $\sum_{i=1}^k\lambda_i$ and settles down in
$\fsplitplus_i(t)$, such that $\dimension\fsplitplus_{i+1}(t)<k\leq
\dimension\fsplitplus_i(t)$.  Formally, these asymptotic embeddings
can be described as:
\begin{equation}
  \label{eq:vol_fmap_fwd}
  \begin{gathered}
    \fpropag(t_1,t)V_k(t_1)
    \underset{t_1\to-\infty}{\subset}\fsplitminus_j(t),\\ 
    \dimension\fsplitminus_{j-1}(t)<k\leq\dimension\fsplitminus_j(t),
  \end{gathered}
\end{equation}
\begin{equation}
  \begin{gathered}
    \label{eq:vol_fmap_bkw}
    \fpropag(t,t_2)^{-1}V_k(t_2)
    \underset{t_2\to+\infty}{\subset}\fsplitplus_j(t),\\
    \dimension\fsplitplus_{j+1}(t)<k\leq\dimension\fsplitplus_j(t).
  \end{gathered}
\end{equation}

Let us now turn to the adjoint propagator $\gpropag(t_1,t_2)$. We
recall that its singular vectors coincide with the singular vectors
for $\fpropag$, while its singular values are reciprocal. Hence the
adjoint Oseledec subspaces can be defined as
\begin{gather}
  \gsplitplus_j(t)=\spanspc
  \left\{
    \fwdvec_{\lambda^{(i)}}(t)\big|i=1,2,\ldots,j
  \right\},\gsplitplus_0(t)=\emptyset,\nonumber \\
  \gsplitplus_1(t)\subset\gsplitplus_2(t)\subset\cdots
  \subset\gsplitplus_s(t)=\mathbb{R}^m,
  \label{eq:gsplit_plus}
\end{gather}
\begin{gather}
  \gsplitminus_j(t)=\spanspc
  \left\{
    \bkwvec_{\lambda^{(i)}}(t)\big|i=j,j+1,\ldots,s
  \right\}, \gsplitminus_{s+1}(t)=\emptyset,\nonumber\\
  \gsplitminus_s(t)\subset\gsplitminus_{s-1}(t)\subset\cdots
  \subset\gsplitminus_1(t)=\mathbb{R}^m.
  \label{eq:gsplit_minus}
\end{gather}
Note that $\gsplitplus_{j-1}(t)\perp \fsplitplus_j(t)$ and
$\gsplitminus_{j+1}(t)\perp \fsplitminus_j(t)$. Reasoning in the same
way as above, we find that the adjoint propagator $\gpropag$
generates the following asymptotic behavior as $t\to\infty$:
\begin{multline}
  \label{eq:asympt_exp_gfwd}
  \vec{v}^{(j)}(t_1)\in
  \gsplitplus_j(t_1)\setminus\gsplitplus_{j-1}(t_1) %
  \Rightarrow \\
  \|\gpropag(t_1,t_1+t)\vec{v}^{(j)}(t_1)\|\sim \myexp{-\lambda^{(j)}t}, 
\end{multline}
\begin{multline}
  \label{eq:asympt_exp_gbkw}
  \vec{v}^{(j)}(t_1)\in
  \gsplitminus_j(t_1)\setminus\gsplitminus_{j+1}(t_1) %
  \Rightarrow \\
  \|\gpropag(t_1-t,t_1)^{-1}\vec{v}^{(j)}(t_1)\|\sim \myexp{\lambda^{(j)}t},
\end{multline}
and the asymptotic embeddings read:
\begin{equation}
  \label{eq:vol_gmap_fwd}
  \begin{gathered}
    \gpropag(t_1,t)V_k(t_1)
    \underset{t_1\to-\infty}{\subset}\gsplitminus_j(t),\\
    \dimension\gsplitminus_{j+1}(t)<k\leq\dimension\gsplitminus_j(t),
  \end{gathered}
\end{equation}
\begin{equation}
  \begin{gathered}
    \label{eq:vol_gmap_bkw}
    \gpropag(t,t_2)^{-1}V_k(t_2)
    \underset{t_2\to+\infty}{\subset}\gsplitplus_j(t),\\
    \dimension\gsplitplus_{j-1}(t)<k\leq\dimension\gsplitplus_j(t).
  \end{gathered}
\end{equation}

\subsection{Finite-time evolution of forward and backward Lyapunov
  vectors}

Now we need to discuss how orthogonal Lyapunov vectors are transformed
in finite time intervals. First consider the action of
$\fpropag(t_1,t_2)$ on forward Lyapunov vectors. For any such vector
related to the $j$th distinct Lyapunov exponent $\lambda^{(j)}$ we
can write: $\fwdvec_{\lambda^{(j)},i}(t_1)\in
\fsplitplus_j(t_1)\setminus\fsplitplus_{j+1}(t_1)$, where
$i=1,2,\ldots,\nu^{(j)}$, see Eq.~\eqref{eq:fsplit_plus}. It means
that this vector shows the asymptotic
behavior~\eqref{eq:asympt_exp_ffwd}, i.e., it grows or decays with the
exponent $\lambda^{(j)}$ forward in time. In turn, it means that
$\fpropag(t_1,t_2)\fwdvec_{\lambda^{(j)},i}(t_1)=\imgfwdvec_{\lambda^{(j)},i}(t_2)
\in \fsplitplus_j(t_2)\setminus\fsplitplus_{j+1}(t_2)$.  We see that
the image of $\fwdvec_{\lambda^{(j)},i}(t_1)$ at $t_2$ is orthogonal
to vectors $\fwdvec_{\lambda^{(n)}}(t_2)$ with $n<j$. But this is not
a forward Lyapunov vector anymore, because the subspaces
$\spanspc\{\fwdvec_{\lambda^{(j)}}(t_2)\}$ and
$\fsplitplus_j(t_2)\setminus\fsplitplus_{j+1}(t_2)$ are not
identical. Vectors from
$\fsplitplus_j(t_2)\setminus\fsplitplus_{j+1}(t_2)$ obligatory have a
nonzero projection inside $\spanspc\{\fwdvec_{\lambda^{(j)}}(t_2)\}$
but typically do not belong to it and also have projections onto
forward vectors with $n>j$.

Let us first assume that there is no degeneracy i.e., all Lyapunov
exponents are distinct. In matrix form we have
$\fpropag(t_1,t_2)\fwdmat(t_1)=\imgfwdmat(t_2)$, where
\begin{equation}\label{eq:fwdmat_def}
  \fwdmat(t)=[\fwdvec_1(t),\fwdvec_2(t),\ldots,\fwdvec_m(t)]
\end{equation}
is the matrix consisting of the forward Lyapunov vectors. According to
the above discussion of $\imgfwdvec_{\lambda^{(j)},i}(t_2)$, the first
vector-column of $\imgfwdmat(t_2)$ is collinear with
$\fwdvec_1(t_2)$. The second one is orthogonal to $\fwdvec_1(t_2)$,
but can have nonzero projections onto all others forward vectors. The
third one is orthogonal both to $\fwdvec_1(t_2)$ and to
$\fwdvec_2(t_2)$ and so on. Thus we can write
\begin{equation}\label{eq:imqfwd_ql_factor}
\imgfwdmat(t_2)=\fwdmat(t_2)\mat{L}, 
\end{equation}
where $\mat{L}$ is a lower triangular matrix. 

When the spectrum of Lyapunov exponents is degenerate, the matrix
$\fwdmat(t_2)$ is not unique. There exist subspaces
$\spanspc\{\fwdvec_{\lambda^{(j)}}(t_2)\}$ corresponding to each
unique Lyapunov exponent, such that any vector from these subspaces
can be treated as a forward Lyapunov vector. This means that the
decomposition~\eqref{eq:imqfwd_ql_factor} is also not unique, because
there exists a variety of non-triangular matrices $\mat L$ satisfying
this equation. But the representation of $\imgfwdmat(t_2)$ as a
product of an orthogonal and a lower triangular matrices exists and is
unique regardless of the degeneracy of Lyapunov exponents. In fact,
this is the well-known QL factorization~\cite{GolubLoan}. The analysis
of the details of the factorization procedure shows that the
orthogonal matrix can always be treated as a matrix of forward
Lyapunov vectors. Hence, regardless of the degeneracy,
Eq.~\eqref{eq:imqfwd_ql_factor} remains valid.

Altogether, the propagator $\fpropag$ maps forward Lyapunov vectors
onto new vectors that are not Lyapunov vectors. In other words,
forward Lyapunov vectors are non-covariant with the dynamics. To
recover forward Lyapunov vectors, we have to perform a QL
factorization. For the subsequent analysis it is convenient to
represent it as a mapping backward in time:
\begin{equation}\label{eq:fmap_fwdvec}
  \fpropag(t_1,t_2)^{-1}\fwdmat(t_2)=\fwdmat(t_1)\mat{L}^\supF(t_1,t_2),
\end{equation}
where $\mat{L}^\supF(t_1,t_2)\in\mathbb{R}^{m\times m}$ is a lower
triangular matrix. Because the propagator is non-singular and QL
factorization is unique (if one requires for all diagonal elements of
$\mat{L}^\supF(t_1,t_2)$ to be positive), this equation determines
$\fwdmat(t_1)$ via $\fwdmat(t_2)$ in a unique way. By definition, the
diagonal elements of $\mat{L}^\supF(t_1,t_2)$ do not vanish, i.e, this
matrix is non-singular.

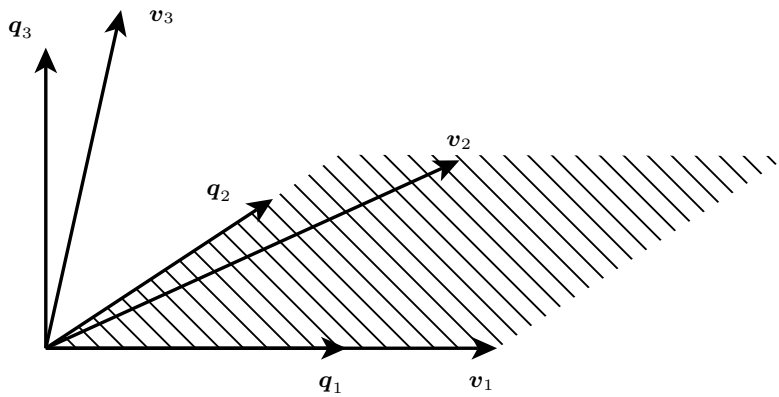
\begin{figure*}
  \centering\ifx\JPicScale\undefined\def\JPicScale{1}\fi
\psset{unit=\JPicScale mm}
\psset{linewidth=0.3,dotsep=1,hatchwidth=0.3,hatchsep=1.5,shadowsize=1,dimen=middle}
\psset{dotsize=0.7 2.5,dotscale=1 1,fillcolor=black}
\psset{arrowsize=1 2,arrowlength=1,arrowinset=0.25,tbarsize=0.7 5,bracketlength=0.15,rbracketlength=0.15}
\begin{pspicture}(0,0)(103.76,55)
\pspolygon[linewidth=0.15,linestyle=none,fillstyle=vlines,hatchwidth=0.25,hatchsep=2](5.25,10.16)(65,10)(103.76,35.58)(44.01,35.74)
\psline[linewidth=0.45,arrowscale=1.6 1.8]{->}(5,10)(65,10)
\psline[linewidth=0.45,arrowscale=1.6 1.8]{->}(5,10)(5,50)
\psline[linewidth=0.45,arrowscale=1.6 1.8]{->}(5,10)(45,10)
\psline[linewidth=0.45,arrowscale=1.6 1.8]{->}(5,10)(35.2,29.86)
\psline[linewidth=0.45,arrowscale=1.6 1.8]{->}(5,10)(60,35)
\psline[linewidth=0.45,arrowscale=1.6 1.8]{->}(5,10)(15,55)
\rput[tl](61.25,6.25){$\vec{v}_1$}
\rput[bl](58.24,36.55){$\vec{v}_2$}
\rput[tl](18.75,55){$\vec{v}_3$}
\rput[tl](41.25,6.25){$\vec{q}_1$}
\rput[br](29.73,29.63){$\vec{q}_2$}
\rput[bl](0,51.25){$\vec{q}_3$}
\end{pspicture}
  \caption{The idea of orthogonalization. The vectors $\vec{v}_i$ are
    the result of mapping~\eqref{eq:propagator_action} and vectors
    $\vec{q}_i$ are their orthogonalization: $\vec{q}_1$ is collinear
    to $\vec{v}_1$, $\vec{q}_2$ belongs to the plane spanned by
    $\vec{v}_1$ and $\vec{v}_2$, and $\vec{q}_3$ belongs to the space
    spanned by vectors $\vec{v}_1$, $\vec{v}_2$, and $\vec{v}_3$.}
  \label{fig:orthog}
\end{figure*}

Repeating the above discussion for the backward Lyapunov vectors, we
see that regardless of the degeneracy, the following relation is
always valid:
\begin{equation}\label{eq:fmap_bkwvec}
  \fpropag(t_1,t_2)\bkwmat(t_1)=\bkwmat(t_2)\mat{R}^\supF(t_1,t_2),
\end{equation}
where 
\begin{equation}\label{eq:bkwmat_def}
  \bkwmat(t)=[\bkwvec_1(t),\bkwvec_2(t),\ldots,\bkwvec_m(t)],
\end{equation}
and $\mat{R}^\supF(t_1,t_2)$ is an upper triangular matrix with a
nonzero diagonal. Like the forward vectors, the backward vectors are
non-covariant with the dynamics.

For the adjoint propagator $\gpropag(t_1,t_2)$ we obtain:
\begin{align}
  \label{eq:gmap_fwdvec}
  \gpropag(t_1,t_2)^{-1}\fwdmat(t_2)&=\fwdmat(t_1)\mat{R}^\supG(t_1,t_2),\\
  \label{eq:gmap_bkwvec}
  \gpropag(t_1,t_2)\bkwmat(t_1)&=\bkwmat(t_2)\mat{L}^\supG(t_1,t_2),
\end{align}
where $\mat{R}^\supG(t_1,t_2)$ and $\mat{L}^\supG(t_1,t_2)$ are upper
and lower non-singular triangular matrices, respectively.

\section{Numerical computation of Lyapunov exponents and 
  forward and backward vectors}\label{sec:num_comp_lyap}

The definition of the Lyapunov exponents and vectors cannot be
implemented directly as a numerical algorithm. It is impossible to
solve Eq.~\eqref{eq:fundam_syst} for a sufficiently long time interval
$t_2-t_1$, to calculate the propagator $\fpropag(t_1,t_2)$, and then
to find a good approximation for the limit matrix $\mat{W}^+$. As we
already discussed above, when we move away from the starting point
$t_1$ almost any vector approaches the first backward Lyapunov vector
$\bkwvec_1(t)$, i.e., falls into subspace $\fsplitminus_1(t)$. Hence,
in this way we can compute only the largest Lyapunov exponent and the
corresponding vector.

Equation~\eqref{eq:fmap_bkwvec} determines a mapping of backward
Lyapunov vectors at $t_1$ onto backward Lyapunov vectors at $t_2$.  A
set of all backward vectors at different times can be considered as a
kind of limit set, attracting or repelling, and the
mapping~\eqref{eq:fmap_bkwvec} can be treated as stationary dynamics
on this set. This gives an idea for an iterative computation of the
backward Lyapunov vectors. One can initialize an arbitrary orthogonal
matrix and start iterations including mapping by $\fpropag$ and QR
factorization as described by Eq.~\eqref{eq:fmap_bkwvec}.  These
iterations converge to the backward Lyapunov vectors where convergence
is guaranteed by Eq.~\eqref{eq:vol_fmap_fwd}. One sees that the
forward in time mapping embeds an arbitrary volume into the subspace
spanned by backward Lyapunov vectors. It means that in the course of
forward iterations $\fpropag(t_n,t_{n+1})\mat{Q}(t_n)=
\mat{Q}(t_{n+1})\mat{R}^\supF(t_{n+1},t_n)$ columns of
$\mat{Q}(t_n)\in\mathbb{R}^{m\times m}$ converge to backward Lyapunov
vectors. In fact this idea was suggested almost simultaneously by
Benettin et al.~\cite{Benettin0,Benettin} and by Shimada and
Nagashima~\cite{Shimada79} to compute the Lyapunov exponents. The
convergence of these iterations towards the backward Lyapunov vectors
is discussed in Refs.~\cite{AGuide, ErshovPotapov98}.

\begin{figure*}
  \centering\ifx\JPicScale\undefined\def\JPicScale{1}\fi
\psset{unit=\JPicScale mm}
\psset{linewidth=0.3,dotsep=1,hatchwidth=0.3,hatchsep=1.5,shadowsize=1,dimen=middle}
\psset{dotsize=0.7 2.5,dotscale=1 1,fillcolor=black}
\psset{arrowsize=1 2,arrowlength=1,arrowinset=0.25,tbarsize=0.7 5,bracketlength=0.15,rbracketlength=0.15}
\begin{pspicture}(0,0)(130.27,58.75)
\pspolygon[linewidth=0.15,fillstyle=vlines,hatchwidth=0.25,hatchsep=2](69.97,10)(115,10)(130.03,30)(85,30)
\pspolygon[linewidth=0.15,fillstyle=vlines,hatchwidth=0.25,hatchsep=2](21.66,50.48)(48.49,37.07)(35.08,10.23)(8.25,23.64)
\psline[linewidth=0.45,arrowscale=1.6 1.8]{->}(70,10)(115,10)
\psline[linewidth=0.45,arrowscale=1.6 1.8]{->}(8.24,23.65)(21.66,50.48)
\psline[linewidth=0.45,arrowscale=1.6 1.8]{->}(8.24,23.65)(35.08,10.23)
\rput{0}(64.19,13.73){\psellipticarc[linewidth=0.7,arrowscale=1.45 2.5]{<-}(0,0)(41.65,41.65){30.71}{129.99}}
\rput[B](66.25,58.75){$\fpropag(t_n,t_{n+1})$}
\rput[br](12.3,39.59){$\vec{q}_2$}
\rput[tr](20.68,13.38){$\vec{q}_1$}
\psline[linestyle=dotted,border=0.3](85,30)(85,10)
\psline[linestyle=dotted,border=0.3](85,30)(70,30)
\psline[linewidth=0.45,arrowscale=1.6 1.8]{->}(70,10)(70,40)
\psline[linewidth=0.45,arrowscale=1.6 1.8]{->}(70,10)(85,30)
\psline(85,11.25)(85,8.75)
\psline[linewidth=0.45,arrowscale=1.6 1.8]{->}(70,10)(100,10)
\psline(71.25,30)(68.75,30)
\psline[linestyle=dotted,border=0.3](115,29.75)(115,10.25)
\rput[b](69.04,42.31){$\vec{q}_2$}
\rput[t](98.23,6.79){$\vec{q}_1$}
\rput[t](84.73,6.79){$r_{12}$}
\rput[tl](109.46,6.79){$\vec{v}_1=r_{11}\vec{q}_1$}
\rput[tl](130.27,6.18){}
\rput[Br](66.34,30.04){$r_{22}$}
\rput[bl](85.07,31.89){$\vec{v}_2$}
\end{pspicture}
  \caption{Computation of a volume after the mapping by
    $\fpropag(t_n,t_{n+1})$.}
  \label{fig:phase_vol_transf_QR}
\end{figure*}
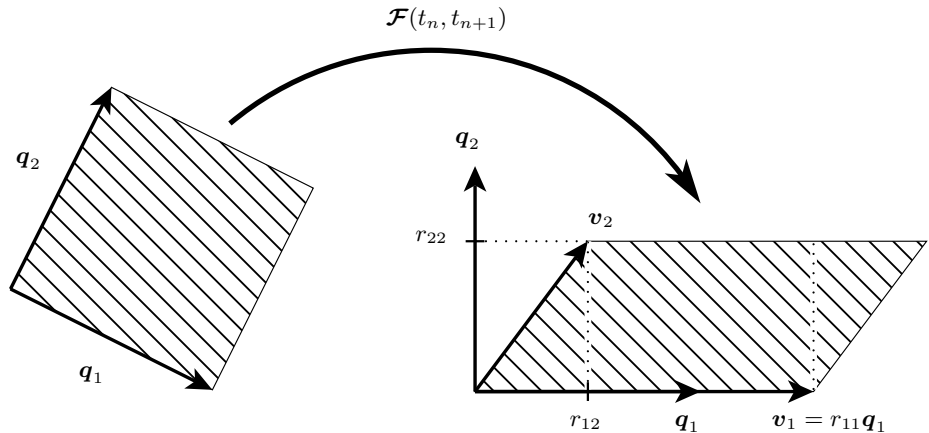

Consider the iterations in more detail, see Fig.~\ref{fig:orthog}.
Suppose we have an orthogonal matrix $\mat{Q}(t_n)$. First we
determine $\fpropag(t_n,t_{n+1})$ for some interval $t_{n+1}-t_n$,
which typically is not very large, and perform the mapping
$\mat{V}(t_{n+1})=\fpropag(t_n,t_{n+1})\mat{Q}(t_n)$. The first
vector-column $\vec{v}_1$ of $\mat{V}(t_{n+1})$ behaves as we need,
namely it approaches the subspace $\fsplitminus_1$.  So, we only
normalize it to prevent overflow or underflow:
$\vec{v}_1\to\vec{q}_1$, $\|\vec{q}_1\|=1$. The plane spanned by
vectors $\vec{v}_1$ and $\vec{v}_2$ approaches the subspace
$\fsplitminus_2$ if $\lambda_1\neq\lambda_2$, or it goes into
$\fsplitminus_1$ otherwise. In the first case we need to prevent the
collapse of the plane due to the alignment of $\vec{v}_2$ along
$\bkwvec_1$, and also the orientation of the plane has to be preserved
to support the convergence. These two goals can be achieved by finding
a new vector $\vec{q}_2$ which is orthogonal to $\vec{q}_1$ and
belongs to the plane originally spanned by $\vec{v}_1$ and
$\vec{v}_2$. This vector is also normalized. In the second case, when
$\lambda_1=\lambda_2$, there is no alignment and, in principle, there
are more options how to define $\vec{q}_2$. But it is allowed anyway
to compute $\vec{q}_2$ as if the degeneracy was absent, and this is
the most reasonable choice making the procedure most transparent.  In
a similar manner we find the third normalized vector $\vec{q}_3$ that
is orthogonal to $\vec{q}_1$ and $\vec{q}_2$ and belongs to the space
spanned by $\vec{v}_1$, $\vec{v}_2$ and $\vec{v}_3$. Doing so for all
the remaining columns of $\mat{V}(t_{n+1})$ we compose the matrix
$\mat{Q}(t_{n+1})$ whose columns are vectors $\vec{q}_i$. Then we use
this $\mat{Q}(t_{n+1})$ as an initial value for the next mapping with
$\fpropag(t_{n+1},t_{n+2})$ and repeat the procedure. After many
recursions the columns of $\mat{Q}(t_n)$ converge to the backward
Lyapunov vectors. This procedure works not only for the whole set of
vectors, but allows to compute any number of the first backward
Lyapunov vectors.

The described procedure eliminates the ambiguity of backward Lyapunov
vectors that emerge when not all Lyapunov exponents are
distinct. Particular directions of backward Lyapunov vectors
corresponding to each degenerate Lyapunov exponent $\lambda^{(j)}$
depend on the choice of the initial matrix $\mat{Q}(t_0)$. But these
variations remain within subspace
$\spanspc\{\bkwvec_{\lambda^{(j)}}\}$ so that any choice is
appropriate. Moreover, in practical computations the degeneracy
manifests itself very weakly, because typically the degenerate
Lyapunov exponents converge to identical values very slowly. In
fact, dealing with a high-dimensional system one needs to know in
advance which of the exponents are expected to be identical to
identify them in the computed spectrum.

The computation of the Lyapunov exponents is illustrated in
Fig.~\ref{fig:phase_vol_transf_QR}. An initial unit square composed of
vectors $\vec{q}_i(t_n)$ is transformed into the parallelogram spanned
by the vectors $\vec{v}_i(t_{n+1})$. After the orthogonalization we
obtain $\vec{q}_i(t_{n+1})$. To compute the area of the parallelogram
we can construct a rectangle with identical area by projecting
$\vec{v}_j(t_{n+1})$ onto $\vec{q}_i(t_{n+1})$:
$r_{ij}=\vec{q}_i\vec{v}_j$.  As we see from the figure, the area is
$r_{11}r_{22}$. Similarly, a $k$-dimensional unit volume after the
mapping is equal to $r_{11}r_{22}\ldots r_{kk}$. Thus, we can define the
local Lyapunov exponents as
\begin{equation}\label{eq:loc_lyap}
  \loclyap_i=\ln (r_{ii})/(t_{n+1}-t_n).
\end{equation}
In the course of the mapping/orthogonalization iterations we need to
accumulate and average $\loclyap_i$ to obtain the Lyapunov exponents.

By construction, the first vector $\vec{v}_1$ has only one nonzero
projection onto $\vec{q}_1$, the second vector $\vec{v}_2$ has two
nonzero projections, onto $\vec{q}_1$ and $\vec{q}_2$, the third
vector $\vec{v}_3$ has three nonzero projections onto first three
vectors $\vec{q}_i$ and so on. It means that $r_{ij}$ are elements of
an upper triangular matrix. So, the procedure described above
represents the matrix $\mat{V}$ as the product
$\mat{V}=\mat{Q}\mat{R}$. Here $\mat{Q}$ is an orthogonal matrix such
that $\spanspc\{\vec{q}_1, \vec{q}_2,\dots
\vec{q}_k\}=\spanspc\{\vec{v}_1, \vec{v}_2,\dots \vec{v}_k\}$ for any
$k\leq m$, and $\mat{R}$ is an upper triangular matrix consisting of
the projections of columns of $\mat{V}$ onto columns of
$\mat{Q}$. This procedure is called QR
factorization~\cite{GolubLoan}. There are different numerical
algorithms of the QR factorization. Note that the often used
Gram--Schmidt algorithm as well as its modified version are not very
accurate when the dimension of the tangent space is
large~\cite{NumLyap10}. Most high precision QR algorithms are based on
so called Householder transformations~\cite{CompMeth,GolubLoan}.

Another way to compute backward Lyapunov vectors is based on the
adjoint propagator $\gpropag$. Equation~\eqref{eq:gmap_bkwvec}
determines the stationary dynamics, and Eq.~\eqref{eq:vol_gmap_fwd}
indicates that the forward iterations converge to this
dynamics. Because $\gpropag(t_n,t_{n+1})$ has reciprocal singular
values, the value $\sigma_m(t_n,t_{n+1})^{-1}$ dominates in the course
of forward iterations with the adjoint propagator. It means that
columns of $\mat{Q}$ converge to the backward Lyapunov vectors in the
reverse order. If we rearrange columns of $\bkwmat$ in
Eq.~\eqref{eq:gmap_bkwvec} in the reverse order, we also have to
transpose $\mat{L}^\supG$ with respect to its diagonal and with
respect to the antidiagonal. As a result we obtain an upper triangular
matrix. Thus, the algorithm is identical to the one previously described. We perform the mapping by $\gpropag(t_n,t_{n+1})$, find a QR
factorization of the resulting matrix, take $\mat{Q}(t_{n+1})$, and do
the next recursion.

Consider now the computation of the forward Lyapunov vectors. The
first algorithm is based on Eqs.~\eqref{eq:vol_gmap_bkw} and
\eqref{eq:gmap_fwdvec}. We need to move backward in time alternating
mappings with $\gpropag(t_n,t_{n+1})^{-1}$ and QR factorizations. The
matrices $\mat{Q}$ converge to $\fwdmat$, and the forward Lyapunov
vectors come up in the correct order. Note that $\gpropag^{-1}$ is
merely the transposition of $\fpropag$, see
Eq.~\eqref{eq:gpropag_defin}. In the course of this procedure we can
compute local Lyapunov exponents as logarithms of diagonal elements of
triangular matrices per unit time. For short time intervals these
local exponents will differ from those given by
Eq.~\eqref{eq:loc_lyap}, but being averaged over many times steps they
also converge to the Lyapunov exponents.

Another algorithm for the forward Lyapunov vectors is based on
Eqs.~\eqref{eq:fmap_fwdvec} and~\eqref{eq:vol_fmap_bkw}. The procedure
is the same as above except using the inverted propagator
$\fpropag^{-1}$. This method computes the vectors in the reversed
order, and, hence, the previous one is usually more applicable. The
idea to apply the transposed propagator instead of the inverted one
was suggested in Ref.~\cite{AGuide}.

The implementation of the algorithm with the transposed propagator
$\fpropag^\supT$ is straightforward for discrete time systems
(e.g. coupled map lattices), where the action of $\fpropag^\supT$ on a
set of (Lyapunov) vectors can be computed using the transposed
Jacobian matrix of the system.  In principle, one can do the same with
continuous systems, but in that case one would have to compute the
full propagator $\fpropag$ first by solving $m$ copies of the
linearized ODE \eqref{eq:basic_ode_lin} and then use its transpose
$\fpropag^\supT$ to evolve the desired number of tangent vectors. This
implementation is inefficient if the system is very high dimensional
($m \gg 1$) and if only a few Lyapunov vectors are to be computed.  As
an alternative, the action of $\fpropag^\supT$ can reformulated as
follows.  Using the Magnus expansion~\cite{Magnus}, we can represent
the propagator of Eq.~\eqref{eq:basic_ode_lin} via matrix exponential
functions as $\fpropag(t_1,t_2)=\exp[\mat\Omega^\supF(t_1,t_2)]$. Here
$\mat\Omega^\supF(t_1,t_2)$ is a matrix that is given as a series
expansion $\mat\Omega^\supF(t_1,t_2)=\sum_{i=1}^\infty
\mat\Omega^\supF_i(t_1,t_2)$, with
$\mat\Omega^\supF_1(t_1,t_2)=\int_{t_1}^{t_2}\mat{J}(\tau_1)d\tau_1$,
$\mat\Omega^\supF_2(t_1,t_2)=\frac{1}{2} \int_{t_1}^{t_2}d\tau_1
\int_{t_1}^{\tau_1}d\tau_2[\mat{J}(\tau_1),\mat{J}(\tau_2)]$, and so
on, see \cite{Magnus}, $\mat{J}(\tau)\equiv \mat{J}(\vec{u},\tau)$ is
the Jacobian matrix, and $[\cdot,\cdot]$ denotes the matrix
commutator. The adjoint propagator reads:
$\gpropag(t_1,t_2)=\fpropag(t_1,t_2)^{-\supT}=
\exp\{-[\mat\Omega^\supF(t_1,t_2)]^\supT\} =
\exp[\mat\Omega^\supG(t_1,t_2)]$.  The matrix
$\mat\Omega^\supG(t_1,t_2)=-[\mat\Omega^\supF(t_1,t_2)]^\supT$
generating $\gpropag(t_1,t_2)$ is obtained with a Magnus expansion
where the Jacobian matrix $\mat{J}(\vec{u},t)$ is replaced by
$-\mat{J}(\vec{u},t)^\supT$. So, to compute the action of
$\gpropag(t_1,t_2)$ on a tangent vector we have to solve the following
linear ODE
\begin{equation}\label{eq:basic_ode_lin_transp}
  \dot{\vec{v}}=-\mat{J}(\vec{u},t)^\supT \vec{v}
\end{equation}
forward in time (from $t_1$ to $t_2 > t_1$, because the action of the
adjoint propagator $\gpropag(t_1,t_2)$ corresponds to moving forward
in time). To compute forward Lypaunov vectors using
$\fpropag(t_1,t_2)^\supT = \gpropag(t_1,t_2)^{-1}$ we have to invert
$\gpropag(t_1,t_2)$. This can be done by integrating the required
number of copies of Eq.~\eqref{eq:basic_ode_lin_transp} and the basic
system \eqref{eq:basic_ode} backward in time (from $t_2$ to $t_1$).

All four algorithms compute the dominating Lyapunov exponents and
corresponding vectors with the highest precision, while the remaining
part of the spectrum is not very accurate. Namely, $\fpropag$- and
$\gpropag^{-1}$-algorithms do the best for the first Lyapunov exponent
and vectors, while $\fpropag^{-1}$- and $\gpropag$-algorithms achieve
the highest accuracy for the $m$th exponent and vectors. One can
perform $\fpropag$- and $\gpropag$-algorithms in parallel, and then
construct weighted sums of computed exponents and backward vectors to
obtain the whole spectrum with very high precision. Similarly,
performing backward iterations simultaneously with $\fpropag^{-1}$ and
$\gpropag^{-1}$ one can compute the forward Lyapunov vectors with
improved accuracy.

\section{Covariant Lyapunov vectors}\label{sec:covar-lyap-vect}

Orthogonal matrices computed according to QR decomposition preserve
subspaces spanned by each \emph{first} $k$ columns of a factorized
matrix. The QL decomposition preserves subspaces spanned by each 
\emph{last} $k$ columns of a factorized matrix. It means that considering
Eqs.~\eqref{eq:fmap_fwdvec} and \eqref{eq:fmap_bkwvec}, we can
conclude, that Oseledec subspaces~\eqref{eq:fsplit_plus}
and~\eqref{eq:fsplit_minus} are preserved under the tangent
flow~\cite{EckRuell85,AGuide}. The same conclusion follows from
Eqs.~\eqref{eq:gmap_fwdvec} and \eqref{eq:gmap_bkwvec} for the
subspaces~\eqref{eq:gsplit_plus} and \eqref{eq:gsplit_minus}:
\begin{equation}
  \begin{gathered}
    \fpropag(t_1,t_2)\fsplitplus_j(t_1) =\fsplitplus_j(t_2),\\
    \fpropag(t_1,t_2)\fsplitminus_j(t_1)=\fsplitminus_j(t_2),
  \end{gathered}
\end{equation}
\begin{equation}
  \begin{gathered}
    \gpropag(t_1,t_2)\gsplitplus_j(t_1) =\gsplitplus_j(t_2),\\
    \gpropag(t_1,t_2)\gsplitminus_j(t_1)=\gsplitminus_j(t_2).
  \end{gathered}
\end{equation}
So, the Oseledec subspaces are invariant under time reversal and
covariant with the dynamics. But this is not the case for the forward
and backward Lyapunov vectors themselves. Being multiplied by
$\fpropag$ and $\gpropag$ they also have to be multiplied by lower or
upper triangular matrices to be mapped to new forward and backward
Lyapunov vectors, see Eqs.~\eqref{eq:fmap_fwdvec},
\eqref{eq:fmap_bkwvec}, \eqref{eq:gmap_fwdvec}, and
\eqref{eq:gmap_bkwvec}.

Given the covariant subspaces, it is natural to search for some
vectors inside these subspaces that are also covariant with the
dynamics and are invariant with respect to time reversal. These
vectors are referred to as \emph{covariant Lyapunov
  vectors}~\cite{GinCLV}. We denote them by $\clvecfwd_j(t)$. The
basic property of these vectors (which are covariant with respect to
the propagator $\fpropag$) can be written as
\begin{equation}\label{eq:clv_main_prop}
  \|\fpropag(t_1,t_1\pm t)\clvecfwd_j(t_1)\|\sim \exp(\pm\lambda_j t)
\end{equation}
for any $t_1$ and $t \to \infty$. The covariant Lyapunov
vectors are norm-independent~\cite{AGuide,WolfCLV}. Also we can
introduce norm-independent adjoint vectors $\clvecadj_j(t)$ that are
covariant with respect to the adjoint dynamics:
\begin{equation}\label{eq:clv_main_prop_adj}
  \|\gpropag(t_1,t_1\pm t)\clvecadj_j(t_1)\|\sim \exp(\mp\lambda_j t).
\end{equation}
Equation~\eqref{eq:clv_main_prop} means that
Eqs.~\eqref{eq:asympt_exp_ffwd} and \eqref{eq:asympt_exp_fbkw} are
fulfilled simultaneously, and Eq.~\eqref{eq:clv_main_prop_adj} implies
the simultaneous validity of Eqs.~\eqref{eq:asympt_exp_gfwd} and
\eqref{eq:asympt_exp_gbkw}. It means that the covariant Lyapunov
vectors belong to the intersection of the Oseledec
subspaces~\cite{Ruelle79,EckRuell85,AGuide}, and the adjoint covariant
vectors can be found within the intersections of the adjoint
subspaces:
\begin{align}
  \label{eq:clv_fsubspc}
  \clvecfwd_j(t)&\in\fsplitplus_j(t)\cap\fsplitminus_j(t),\\
  \label{eq:clv_gsubscp}
  \clvecadj_j(t)&\in\gsplitplus_j(t)\cap\gsplitminus_j(t).
\end{align}
These intersections are always nonempty because the sum of dimensions
of Oseledec subspaces is always higher than the dimension of the whole
tangent space. 

Consider arbitrary vectors $\vec{v}^{(j)}(t_1)\in
\fsplitplus_j(t_1)\setminus\fsplitplus_{j+1}(t_1)$, where
$j=1,2,\ldots, s$, and $s$ is the number of distinct Lyapunov
exponents. There are $\nu^{(j)}$ linearly independent vectors
corresponding to the $j$th Lyapunov exponent $\lambda^{(j)}$, and the
total number of such vectors is
$\sum_{j=1}^s\nu^{(j)}=m$. Representing the whole set of these vectors
as a matrix $\mat{V}$, we obtain $\mat{V}=\fwdmat\mat{A}^+$, where
$\mat{A}^+$ is a lower triangular matrix, and $\fwdmat$ is a matrix of
forward Lyapunov vectors~\eqref{eq:fwdmat_def}. As follows from
Eq.~\eqref{eq:asympt_exp_ffwd}, when the forward propagator $\fpropag$
is applied to these vectors, the first $\nu^{(1)}$ of them grow or
decay asymptotically with the exponent $\lambda^{(1)}$, the next
$\nu^{(2)}$ vectors grow / decay with the exponent $\lambda^{(2)}$ and
so on. In a similar manner we can consider arbitrary vectors
$\vec{v}^{(j)}(t_1)\in
\fsplitminus_j(t_1)\setminus\fsplitminus_{j-1}(t_1)$. The matrix of
these vectors $\mat{V}$ can be found as $\mat{V}=\bkwmat\mat{A}^-$,
where $\mat{A}^-$ is an upper triangular matrix, and $\bkwmat$ is
defined by Eq.~\eqref{eq:bkwmat_def}. According to
Eq.~\eqref{eq:asympt_exp_fbkw}, acting upon these vectors by the
inverted propagator $\fpropag^{-1}$, we can observe that the first
$\nu^{(1)}$ of them grow or decay asymptotically with the exponent
$-\lambda^{(1)}$, the next $\nu^{(2)}$ vectors grow / decay with the
exponent $-\lambda^{(2)}$ and so on. Let
$\clmatfwd(t)=[\clvecfwd_1(t),\clvecfwd_2(t),\ldots,\clvecfwd_m]$ be a
matrix consisting of the covariant Lyapunov vectors, and let
$\clmatadj(t)=[\clvecadj_1(t),\clvecadj_2(t),\ldots,\clvecadj_m]$ be a
matrix of adjoint covariant vectors. As follows from
Eq.~\eqref{eq:clv_main_prop}, the covariant vectors have to
demonstrate both forward~\eqref{eq:asympt_exp_ffwd} and
backward~\eqref{eq:asympt_exp_fbkw} asymptotic behavior. It means that
there exist an upper triangular matrix $\mat{A}^-$ and a lower
triangular matrix $\mat{A}^+$, such that
\begin{equation}\label{eq:clv_defin}
  \clmatfwd(t)=\bkwmat(t)\mat{A}^-(t)=\fwdmat(t)\mat{A}^+(t).
\end{equation}
Reasoning in a similar manner one obtains
for the adjoint vectors:
\begin{equation}\label{eq:clv_defin_adj}
  \clmatadj(t)=\fwdmat(t)\mat{B}^+(t)=\bkwmat(t)\mat{B}^-(t),
\end{equation}
where $\mat{B}^+(t)$ and $\mat{B}^-(t)$ are upper and lower triangular
matrices, respectively. Note that Eqs.~\eqref{eq:clv_defin} and
\eqref{eq:clv_defin_adj} convey, in fact, the same property of
covariant vectors as Eq.~\eqref{eq:clv_fsubspc} and
\eqref{eq:clv_gsubscp}, respectively. Multiplying
Eq.~\eqref{eq:clv_defin} by $[\fwdmat(t)]^\supT$ and
Eq.~\eqref{eq:clv_defin_adj} by $[\bkwmat(t)]^\supT$ we obtain the
relations between triangular matrices that will be required later:
\begin{gather}
  \label{eq:lu_factor}
  \mat{P}(t)\mat{A}^-(t)=\mat{A}^+(t),\\
  \label{eq:lu_factor_adj}
  \mat{P}(t)^\supT\mat{B}^+(t)=\mat{B}^-(t),
\end{gather}
where 
\begin{equation}\label{eq:def_mat_p}
  \mat{P}(t)=[\fwdmat(t)]^\supT\bkwmat(t)
\end{equation}
is a $m\times m$ orthogonal matrix.

If the Lyapunov exponents are degenerate, the covariant vectors are
not unique. Let us discuss what Eq.~\eqref{eq:clv_defin} implies in
this case (Eq.~\eqref{eq:clv_defin_adj} can be considered in the same
way).  If $\clmatfwd(t)$ is known, then we can compute $\bkwmat(t)$
and $\mat{A}^-(t)$, and $\fwdmat(t)$ and $\mat{A}^+(t)$ via QR and QL
decompositions, respectively, in a unique way.  However,
Eq.~\eqref{eq:clv_defin} does not determine $\clmatfwd(t)$ via
$\fwdmat(t)$ and $\bkwmat(t)$ in a unique way. In principle, there
exist orthogonal matrices $\bkwmat$ and $\fwdmat$ that allow one to
fulfill Eq.~\eqref{eq:clv_defin} with several couples $\mat{A}^-(t)$
and $\mat{A}^+(t)$, resulting in different matrices $\clmatfwd$, and,
hence, in different covariant Lyapunov vectors. As an example one can
consider a matrix $\fwdmat$ that consists of columns of $\bkwmat$
arranged in the reverse order. Ambiguity of $\mat{A}^-(t)$ and
$\mat{A}^+(t)$ means that there are Lyapunov exponents associated with
several covariant vectors. But on the other hand, the total number of
covariant vectors is equal to the total number of Lyapunov exponents
$m$, and there are no exponents without vectors. It means that the
ambiguity can occur if and only if the Lyapunov exponents are
degenerate. The covariant vectors associated with a $k$ times
degenerate Lyapunov exponent can have arbitrary orientation within a
$k$-dimensional subspace corresponding to this exponent. But because
any set of linearly independent covariant vectors from the subspace
corresponding to the degenerate exponent is as good as any other, this
ambiguity can be ignored: we just need to have any linear independent
set of vectors. (We recall that though forward and backward vectors
are also subject to the degeneracy, their ambiguity is eliminated in
the course of the computations, see Sec.~\ref{sec:num_comp_lyap}.)

Let us find how $\clmatfwd(t_1)$ is transformed by
$\fpropag(t_1,t_2)$. In general we can write
\begin{equation}\label{eq:clv_fmap}
  \fpropag(t_1,t_2)\clmatfwd(t_1)=\clmatfwd(t_2)\matcfwd(t_1,t_2),
\end{equation}
where $\matcfwd(t_1,t_2)$ is a matrix whose structure should be
determined. When the Lyapunov spectrum is not degenerate,
Eq.~\eqref{eq:clv_fsubspc} immediately implies that $\matcfwd(t_1,t_2)$
is diagonal. To show that this is the case regardless of the
degeneracy, we substitute $\clmatfwd(t)=\fwdmat(t)\mat{A}^+(t)$, see
Eq.~\eqref{eq:clv_defin}, in Eq.~\eqref{eq:clv_fmap} and, taking into
account Eq.~\eqref{eq:fmap_fwdvec}, we obtain
\begin{equation}\label{eq:lower_triang_map}
  \mat{L}^\supF(t_1,t_2)\mat{A}^+(t_2)\matcfwd(t_1,t_2)=\mat{A}^+(t_1).
\end{equation}
Since all known matrices here are lower triangular, $\matcfwd(t_1,t_2)$
is also lower triangular. Analogously substituting
$\clmatfwd(t)=\bkwmat(t)\mat{A}^-(t)$ in  Eq.~\eqref{eq:clv_fmap} and
using Eq.~\eqref{eq:fmap_bkwvec} we obtain:
\begin{equation}\label{eq:upper_triang_map}
  \mat{R}^\supF(t_1,t_2)\mat{A}^-(t_1)=\mat{A}^-(t_2)\matcfwd(t_1,t_2),
\end{equation}
i.e., $\matcfwd(t_1,t_2)$ is an upper triangular matrix. Simultaneous
upper and lower triangular structure has only a diagonal matrix:
$\matcfwd(t_1,t_2)=\diagmat[c_1(t_1,t_2),c_2(t_1,t_2),\ldots,c_m(t_1,t_2)]$.
Hence, the vectors $\clvecfwd_j$ can freely evolve under the tangent
flow~\eqref{eq:clv_fmap} so that the tangent flow preserves their
directions. The direction, represented by $\clvecfwd_j(t_1)$ at $t_1$
is mapped onto the direction pointed by $\clvecfwd_j(t_2)$ at $t_2$,
and the backward step maps $\clvecfwd_j(t_2)$ onto the direction of
$\clvecfwd_j(t_1)$. The vectors themselves are stretched or contracted
by factors $\matcfwdelem_i(t_1,t_2)$. (Recall, that the directions of the forward
and the backward Lyapunov vectors are not preserved). The adjoint
vectors freely evolve under the tangent flow generated by the adjoint
propagator:
\begin{equation}
  \label{eq:clv_gmap}
  \gpropag(t_1,t_2)\clmatadj(t_1)=\clmatadj(t_2)[\matcadj(t_1,t_2)]^{-1}.
\end{equation}
One can say that the vectors $\clvecfwd_j$ are covariant with the
tangent dynamics generated by $\fpropag$ and the adjoint vectors
$\clvecadj_j$ are covariant with the tangent dynamics of
$\gpropag$. This is the reason why these vectors are referred to as
covariant vectors.

Since the covariant vectors are defined up to an arbitrary length, the
diagonal elements of $\matcfwd$ can be defined in various ways. In
particular, to fulfill Eq.~\eqref{eq:clv_main_prop} we should not
normalize the vectors, and $\matcfwd\equiv\identmat$ in this
case. However, in the course of numerical computations we need to
avoid overflows and underflows.  Hence, constant lengths of
$\clvecfwd_j(t)$ have to preserved with respect to the chosen norm. In
this case
$\matcfwdelem_j(t_1,t_2)=||\fpropag(t_1,t_2)\clvecfwd_j(t_1)||/||\clvecfwd_j(t_1)||$,
and
\begin{equation}\label{eq:norm_dep_loc_lyap}
  \ln[\matcfwdelem_j(t_1,t_2)]/(t_2-t_1)
\end{equation}
can be treated as local Lyapunov exponent. The values of these local
Lyapunov exponents depend on the norm, but being averaged over many /
long time intervals $(t_1,t_2)$, regardless of the norm they converge
to the Lyapunov exponents $\lambda_j$. Consider an important
particular case. As follows from the discussions in
Sec.~\ref{sec:num_comp_lyap}, one can build unit volumes using the
covariant Lyapunov vectors when the diagonal elements of the upper
triangular matrix $\mat{A}^-$ are equal to 1, see
Eq.~\eqref{eq:clv_defin}. Equation~\eqref{eq:upper_triang_map}
describes the dynamics of $\mat{A}^-$ corresponding to the tangent
dynamics of the covariant Lyapunov vectors. When two upper triangular
matrices are multiplied, the resulting matrix is also upper triangular
and its diagonal elements are the products of the diagonal elements of
the multipliers. Thus, if the covariant Lyapunov vectors are rescaled
to preserve ones on the diagonal of $\mat{A}^-$, then the $\matcfwdelem_j$ are
equal to the diagonal elements of $\mat{R}^\supF$, and the local
Lyapunov exponents \eqref{eq:norm_dep_loc_lyap} coincide with those
defined by Eq.~\eqref{eq:loc_lyap}:
$\ln[\matcfwdelem_j(t_1,t_2)]/(t_2-t_1)=\loclyap_j(t_1,t_2)$.

Let us now discuss what it means if covariant vectors merge.  The
phase space of dynamical systems can contain structures called ``wild
hyperbolic sets'' that are responsible for the existence of
structurally stable and unavoidable homoclinic tangencies between
stable and unstable manifolds. In turn, the presence of these
tangencies results in formation of non-hyperbolic chaotic
attractors~\cite{GuckHolm83}. Since covariant vectors are associated
with invariant manifolds of trajectories, in points of tangencies the
corresponding vectors become
collinear~\cite{EckRuell85,GinCLV,WolfCLV,EffDim}. The same happens
with the corresponding adjoint covariant vectors. Collinear vectors
result in a singularity of the matrices $\clmatfwd(t)$, and
$\clmatadj(t)$. The triangular matrices $\mat{A}^\pm(t)$ and
$\mat{B}^\pm(t)$ also become singular. Note, that this property is
time-invariant: as follows from Eq.~\eqref{eq:clv_fmap} and
\eqref{eq:clv_gmap} if some of covariant vectors are identical at
$t=t_1$, they remain identical for all time. In practice, selecting an
arbitrary trajectory we almost never hit exactly the trajectory with
the tangencies. But if a trajectory with tangencies exists, the
arbitrarily selected orbit will pass infinitely close to it and we
will encounter with a nonzero frequency ill-conditioned matrices of
covariant vectors. Note that this is not the case for orthogonal
forward and backward vectors, which are not affected by tangencies.

Now we consider how covariant and adjoint covariant vectors are
related to each other. First of all notice that given $\clmatfwd(t)$,
one can always compute $\bkwmat(t)$ and $\mat{A}^-(t)$ as its QR
decomposition and $\fwdmat(t)$ and $\mat{A}^+(t)$ as a QL
decomposition. Then one can construct the matrix
$\mat{P}(t)=[\fwdmat(t)]^\supT\bkwmat(t)$ and compute $\clmatadj(t)$
via the LU method as described below in
Sec.~\ref{sec:lu-factorization}. It means that these two sets of
vectors are not independent from each other. However, the mutual
orientation of these vectors can help to recover some new
data. 

Transposing Eq.~\eqref{eq:clv_defin_adj} and multiplying it with
Eq.~\eqref{eq:clv_defin}, we obtain:
$\mat{B}^+(t)^\supT\mat{A}^+(t)=\mat{B}^-(t)^\supT\mat{A}^-(t)$.  The
left hand side of this equation is a lower triangular matrix, while
the matrix on the right hand side is upper triangular. Hence,
\begin{equation}\label{eq:signat_triang}
  \mat{B}^\pm(t)^\supT\mat{A}^\pm(t)=
  \mat{A}^\pm(t)^\supT\mat{B}^\pm(t)=\signat(t), 
\end{equation}
where $\signat(t)$ is a diagonal matrix. Again take into account
Eqs.~\eqref{eq:clv_defin_adj} and \eqref{eq:clv_defin} to write:
\begin{equation}\label{eq:signat}
  \clmatadj(t)^\supT\clmatfwd(t)=
  \clmatfwd(t)^\supT\clmatadj(t)=\signat(t).
\end{equation}
The diagonal structure of $\signat$ indicates that each adjoint
covariant vector $\clvecadj_j(t)$, $j=1,2,\ldots,m$ is always
orthogonal to the covariant vectors $\clvecfwd_i(t)$, where $i\neq
j$. In presence of the tangency $\clvecfwd_j(t)=\clvecfwd_{j+1}(t)$
the $j$th and the $(j+1)$th diagonal elements of $\mat{D}$ vanish,
i.e., corresponding adjoint and original vectors also become
orthogonal: $\clvecfwd_{j+i}(t)\perp \clvecadj_{j+i}(t)$, where
$i=0,1$.  It means that given the vectors $\clvecfwd_i(t)$, one can
find the adjoint vectors $\clvecadj_j(t)$ as null vectors of the
matrix consisting of all $\clvecfwd_i(t)$ except the $j$th one. Notice
that even if a tangency occurs, one still can compute $\clvecadj_j(t)$
in this way. To find how $\signat(t)$ is varying in time, we transpose
Eq.~\eqref{eq:clv_fmap}, multiply it with Eq.~\eqref{eq:clv_gmap}, and
take into account Eq.~\eqref{eq:gpropag_defin}:
$\clmatfwd(t_1)^\supT\clmatadj(t_1)=
\matcfwd(t_1,t_2)\clmatfwd(t_2)^\supT\clmatadj(t_2)[\matcadj(t_1,t_2)]^{-1}$. Hence,
$\signat(t_1)=\matcfwd(t_1,t_2) \signat(t_2)[\matcadj(t_1,t_2)]^{-1}$
(recall that all matrices here are diagonal). Altogether, the elements
of the diagonal matrix $\mat{D}$ are cosines of angles between
corresponding covariant and adjoint covariant vectors. Since these
angles are affected by tangencies, their time averages as well as
their temporal fluctuations; i.e., the first and other moments, can be
considered as characteristic numbers describing the structure of an
attractor. The angles are norm-independent, because they are defined
in terms of covariant and adjoint covariant Lyapunov vectors which
share this property.

If the covariant vectors are computed with a non-ideal accuracy, the
errors will grows in course of the tangent dynamics.  The same is the
case for the adjoint covariant vectors. In particular, it means that
if we have found numerically covariant vectors at $t_1$, we cannot
compute them at $t>t_1$ via Eq.~\eqref{eq:clv_fmap} because numerical
errors results in the divergence from the true directions. But
nevertheless, Paz\'o in Ref.~\cite{Pazo2010} shows that this
divergence is actually sufficiently slow. Hence,
Eq.~\eqref{eq:clv_fmap} can be used to find an estimate for the
covariant vectors at $t>t_1$ when $t-t_1$ is not very large.

The covariant Lyapunov vectors are defined locally, according
to Eqs.~\eqref{eq:clv_defin} and \eqref{eq:clv_defin_adj}, and
asymptotically, as follows from Eqs.~\eqref{eq:clv_main_prop} and
\eqref{eq:clv_main_prop_adj}. These equations provide two basic ideas
for computing these vectors. The first one is to find backward and
forward Lyapunov vectors for some point of the trajectory and compute
an intersection of corresponding Oseledec subspaces. The
straightforward implementation of this approach though possible, takes
a lot of computational resources. We discuss it in
Sec.~\ref{sec:inters-osel-subsp}. In Secs.~\ref{sec:lu-factorization}
and \ref{sec:compl-subsp} more ``clever'' implementations are
considered.

The second approach is to try to arrive at asymptotic behavior
described by Eq.~\eqref{eq:clv_main_prop} or
\eqref{eq:clv_main_prop_adj}. If we initialize a vector, satisfying
Eq.~\eqref{eq:asympt_exp_fbkw} and start iterations backward in time,
after a long time we closely approach the limiting vectors that evolve
as $\fpropag(t_n,t_{n+1})^{-1}\vec{v}_j(t_{n+1})=
\vec{v}_j(t_n)c_j(t_n,t_{n+1})^{-1}$, where $c_j(t_n,t_{n+1})$ are
related to the local Lyapunov exponents
\eqref{eq:norm_dep_loc_lyap}. This equation is reversible, so that
when the limit is reached, we can turn forward and arrive the opposite
limit too. It means that the limiting vectors $\vec{v}_j$ found in
this way satisfy Eq.~\eqref{eq:clv_main_prop} and coincide with
$\clvecfwd_j$. The forward iterations defined by
Eqs.~\eqref{eq:asympt_exp_ffwd} also converge to the covariant
Lyapunov vectors. Similarly, the iterations initialized according to
Eqs.~\eqref{eq:asympt_exp_gfwd} and \eqref{eq:asympt_exp_gbkw}
converge to the adjoint covariant vectors. The straightforward
numerical implementation of this approach is impossible. Due to
numerical noise, vectors $\vec{v}^{(j)}$ cannot be initialized
exactly as required, and the numerical routines always converge to the
single dominating vector. But a way to avoid this obstacle is known,
and we consider it in Sec.~\ref{sec:iterations}.

\section{Numerical methods for computing covariant Lyapunov
  vectors}\label{sec:pract-meth-comp}

\subsection{Intersection of Oseledec subspaces}\label{sec:inters-osel-subsp}

A straightforward way to find covariant Lyapunov vectors is based on
Eq.~\eqref{eq:clv_fsubspc}. Given forward and backward Lyapunov
vectors, one can construct intersections of the Oseledec subspaces and
find the covariant vectors. To compute the intersection of two
subspaces one can compute so called principle angles between
subspaces~\cite{GolubLoan,KnyArg02}. In brief, this method is
associated with computation of the singular values and vectors of
submatrices of the matrix~\eqref{eq:def_mat_p}.

To compute the $j$th covariant vector one needs the first $j$
backward vectors and $m-j+1$ last forward vectors. The first backward
vectors can be computed in the course of the iterations with the
propagator $\fpropag$, and the last forward vectors are the result of
the iterations with the inverted propagator $\fpropag^{-1}$, see
Sec.~\ref{sec:num_comp_lyap}.

Regardless of $j$, $m+1$ forward and backward Lyapunov vectors are
always required. So, this method is applicable for computation of the
whole spectrum, but this is not an efficient approach if one needs
only a few first covariant vectors. Because the forward Lyapunov
vectors are computed in the reverse order, this method has a
``flattened'' accuracy along the spectrum: the backward vectors have
higher accuracy in first part of the spectrum, and the forward one are
more accurate in its last part. So, the resulting covariant vectors
have approximately the same accuracy for the whole spectrum.

\subsection{Method of LU factorization}\label{sec:lu-factorization}

It is possible to avoid computation of the whole spectrum of the
forward or backward Lyapunov vectors to get only a few first covariant
vectors. Two original ideas, which were reported in
Refs.~\cite{WolfCLV,GinCLV}, are discussed in
Secs.~\ref{sec:compl-subsp} and \ref{sec:iterations}. In the current
section we present a new approach to this problem.

Consider Eq.~\eqref{eq:lu_factor}. Matrices $\mat{A}^+$ and
$\mat{A}^-$ are lower and upper triangular, respectively. If
$\mat{A}^-$ is non-singular, we can rewrite Eq.~\eqref{eq:lu_factor} as
$\mat{P}=\mat{A}^+(\mat{A}^-)^{-1}$. This equation can be considered
as an LU factorization of $\mat{P}$, i.e., representation of a matrix
as a product of a lower and an upper triangular
matrix~\cite{GolubLoan}. If the factorization exists, it is unique
up to the diagonal elements of one of the matrices (factors). For us
it means that if we find the LU decomposition of $\mat{P}$, we find
the covariant vectors up to arbitrary lengths.

There are many well developed standard routines computing the LU
factorization.  But for us the serious disadvantage is that they work
well only as long as the assumption of non-singularity of $\mat{A}^-$
remains valid. If matrices $\mat{A}^\pm$ are singular, the
straightforward factorization of $\mat{P}$ does not exist. The
standard routines for LU decomposition avoid this obstacle performing
preliminary permutations of rows and columns of $\mat{P}$. This is not
suitable for us, because the order of rows and columns in $\mat{P}$ is
essential. Moreover, the standard routines find both $\mat{A}^-$, and
$\mat{A}^+$, while it is enough for us to have only $\mat{A}^-$.

Let us return to Eq.~\eqref{eq:lu_factor}. We shall demonstrate now
that the required elements of $\mat{A}^-$ can be found from this
equation regardless of a possible singularity of $\mat{A}^\pm$. To
compute the $j$th covariant vector we need to find the top $j$
elements of the $j$th column of $\mat{A}^-$. This fragment of the
column can be denoted as $\mat{A}^-(1\mycolon j, j)$. The remaining
fragment $\mat{A}^-(j+1\mycolon m, j)$ contains zeros. Note that here
we omit the time dependence and use parentheses to indicate
submatrices. The matrix equation for nonzero elements reads:
$\mat{P}(1\mycolon j,1\mycolon j)\mat{A}^-(1\mycolon
j,j)=\mat{A}^+(1\mycolon j,j)$, where $\mat{P}(1\mycolon j,1\mycolon
j)$ is the top left square submatrix of $\mat{P}$. Because $\mat{A}^+$
is lower triangular, the fragment $\mat{A}^+(1\mycolon j,j)$ of its
$j$th column contains zeros except for the diagonal element
$\mat{A}^+(j,j)$. As already mentioned above, the LU decomposition is
unique up to diagonal elements of one of the matrices. It means that
we can eliminate the equation, corresponding to the $j$th row of
$\mat{P}(1\mycolon j,1\mycolon j)$ and write the following homogeneous
matrix equation
\begin{equation}\label{eq:lu_find_clv}
  \mat{P}(1\mycolon j-1,1\mycolon j)\mat{A}^-(1\mycolon j,j)=\vec{0}.
\end{equation}
This equation allows to compute nonzero elements of the $j$th column
of $\mat{A}^-$ as the null space of the rectangular submatrix
$\mat{P}(1\mycolon j-1,1\mycolon j)$. To obtain covariant unit vectors
the solutions have to be normalized.

Equation~\eqref{eq:lu_find_clv} can, in principle, have multiple
solutions for $\mat{A}^-(1\mycolon j,j)$. (In this case the rank of
$\mat{P}(1\mycolon j-1,1\mycolon j)$ is less than $(j-1)$.)  As we
discussed above, this ambiguity can occur only due to the degeneracy
of the Lyapunov exponents, and we can arbitrarily choose one of the
multiple solutions.

As follows from Eq.~\eqref{eq:lu_factor_adj}, the adjoint covariant
vectors can be computed analogously, using the equation
\begin{equation}\label{eq:lu_find_clv_adj}
  (\mat{P}^\supT)(1\mycolon j-1,1\mycolon j)\mat{B}^+(1\mycolon j,j)=\vec{0}.
\end{equation}

Let us now consider the submatrix $\mat{P}(1\mycolon j,1\mycolon j)$.
If this is singular, then Eq.~\eqref{eq:lu_find_clv} provides for the
$(j+1)$th column the solution $\mat{A}^-(1\mycolon
m,j+1)=\mat{A}^-(1\mycolon m,j)$, i.e., the $j$th and $(j+1)$th
covariant vectors coincide. The inverse is also true, and, hence, the
singularity of the submatrix $\mat{P}(1\mycolon j,1\mycolon
j)$ is a sufficient and necessary condition for merging of the $j$th
and $(j+1)$th covariant vectors.

As discussed above, the merging of covariant Lyapunov vectors
indicates tangencies of invariant manifolds of an attractor that, in
particular, occur when the attractor is chaotic and
non-hyperbolic~\cite{GuckHolm83}. To detect the violation of
hyperbolicity, one usually studies a distribution of angles between
expanding and contracting subspaces spanned by corresponding covariant
vectors~\cite{NonhypTan93, Anishchenko2000301, GinCLV,
  HyperSpace08}. (Another method for a numerical test of
hyperbolicity, which does not employ covariant vectors, is based on
the so called cone criterion~\cite{KuzSat07}.)  Analyzing properties
of submatrices of $\mat{P}$ one can test for hyperbolicity without
explicit computation of covariant vectors. Let the number of positive
Lyapunov exponents be $k$.  Moving along a trajectory, we need to
compute some characteristic number whose small value indicates the
nearness of $\mat{P}(1\mycolon k,1\mycolon k)$ to singularity. It can
be, for instance, the determinant or the smallest singular value. A
small characteristic number means that the trajectory passes close to
the tangency. So, if the distribution of characteristic numbers
computed for many trajectory points is well separated from the origin,
then the chaos is hyperbolic, and if it approaches the origin
violations of hyperbolicity occur.

One can also study the statistics of nearness to singularity of all
submatrices $\mat{P}(1\mycolon j,1\mycolon j)$, where $j=1,2,\ldots,
m-1$. This can provide detailed information concerning properties of
various limit sets embedded in an attractor.

Another way to characterize an attractor is to compute the matrix
$\signat$ containing cosines of angles between covariant and adjoint
covariant vectors. As discussed above, each merged couple of vectors,
i.e., each tangency, is represented as a couple of zeros of the
corresponding matrix elements. To compute $\signat$, first we find the
matrix $\mat{A}^-$, then using Eq.~\eqref{eq:lu_factor} compute only
the diagonal elements of $\mat{A}^+$, and after that compute
$\mat{B}^+$ using Eq.~\eqref{eq:lu_find_clv_adj}. (Though only its
diagonal elements are required, we cannot get them without computing
the rests of the columns.)  Finally, we obtain the elements of
$\signat$ as products of diagonal elements of $\mat{A}^+$ and
$\mat{B}^+$; see Eq.~\eqref{eq:signat_triang}. Note, that it is not
required to compute the whole matrix $\signat$. The method allows one
to find only a few first elements.

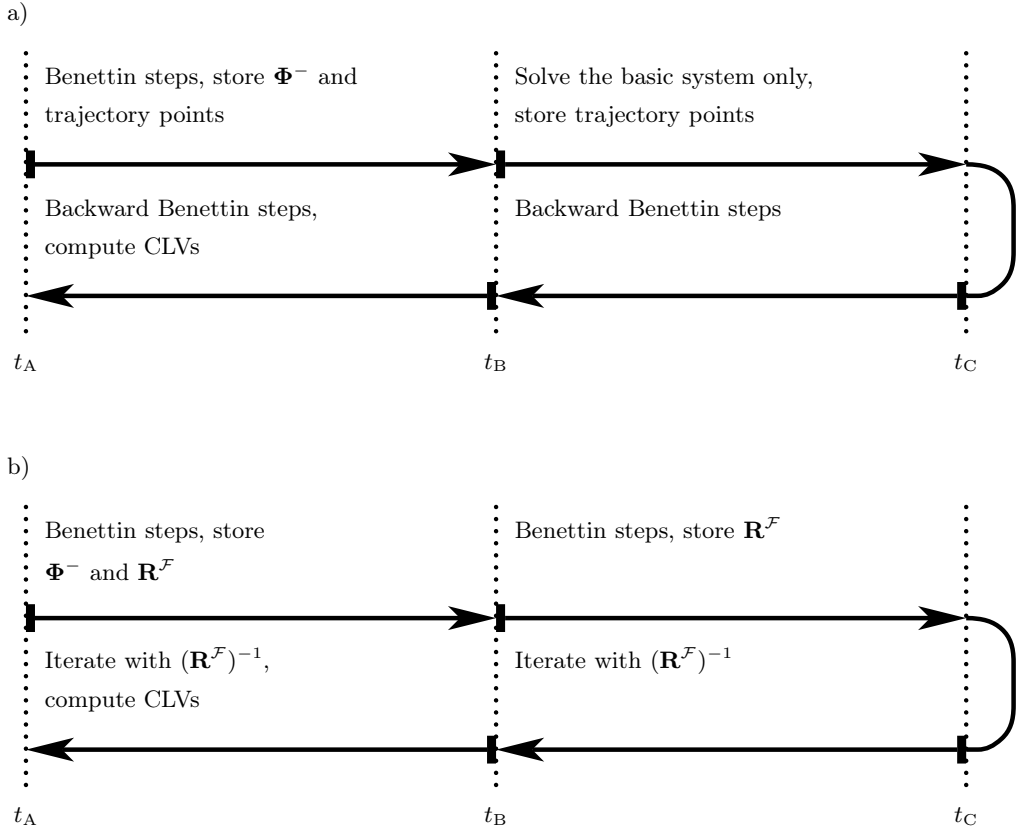
\begin{figure*}
  \begin{center}
    \ifx\JPicScale\undefined\def\JPicScale{1}\fi
\psset{unit=\JPicScale mm}
\psset{linewidth=0.3,dotsep=1,hatchwidth=0.3,hatchsep=1.5,shadowsize=1,dimen=middle}
\psset{dotsize=0.7 2.5,dotscale=1 1,fillcolor=black}
\psset{arrowsize=1 2,arrowlength=1,arrowinset=0.25,tbarsize=0.7 5,bracketlength=0.15,rbracketlength=0.15}
\begin{pspicture}(0,0)(133.8,50)
\psline[linewidth=0.6,linestyle=dotted](2.5,7.5)(2.5,45)
\psline[linewidth=0.6,linestyle=dotted](65,7.5)(65,45)
\psline[linewidth=0.6,linestyle=dotted](127.5,7.5)(127.5,45)
\psline[linewidth=0.6,arrowscale=1 2,arrowsize=2 2]{|->}(2.5,30)(65,30)
\psline[linewidth=0.6,arrowscale=1 2,arrowsize=2 2]{|->}(65,30)(127.5,30)
\pscustom[linewidth=0.6]{\psbezier(127.5,30)(129.7,30)(131.12,29.56)(132.23,28.54)
\psbezier(133.33,27.53)(133.8,26.22)(133.8,24.18)
\psbezier(133.8,22.13)(133.8,20.39)(133.8,18.34)
\psbezier(133.8,16.3)(133.33,14.99)(132.23,13.97)
\psbezier(131.12,12.95)(130.18,12.51)(129.08,12.51)
\psbezier(127.97,12.51)(127.5,12.51)(127.5,12.51)
\psbezier(127.5,12.51)(127.5,12.51)(127.5,12.51)
\psbezier(127.5,12.51)(127.5,12.51)(127.5,12.51)
}
\psline[linewidth=0.6,arrowscale=1 2,arrowsize=2 2]{|->}(127.5,12.5)(65,12.5)
\psline[linewidth=0.6,arrowscale=1 2,arrowsize=2 2]{|->}(65,12.5)(2.5,12.5)
\rput[bl](5,40){Benettin steps, store $\bkwmat$ and}
\rput[bl](5,35){trajectory points}
\rput[bl](67.5,40){Solve the basic system only,}
\rput[bl](67.5,35){store trajectory points}
\rput[bl](67.5,22.5){Backward Benettin steps}
\rput[bl](5,22.5){Backward Benettin steps,}
\rput[bl](5,17.5){compute CLVs}
\rput[b](2.5,2.5){$t_\text{A}$}
\rput[b](65,2.5){$t_\text{B}$}
\rput[b](127.5,2.5){$t_\text{C}$}
\rput[l](0,50){a)}
\end{pspicture}\\[1cm]
    \ifx\JPicScale\undefined\def\JPicScale{1}\fi
\psset{unit=\JPicScale mm}
\psset{linewidth=0.3,dotsep=1,hatchwidth=0.3,hatchsep=1.5,shadowsize=1,dimen=middle}
\psset{dotsize=0.7 2.5,dotscale=1 1,fillcolor=black}
\psset{arrowsize=1 2,arrowlength=1,arrowinset=0.25,tbarsize=0.7 5,bracketlength=0.15,rbracketlength=0.15}
\begin{pspicture}(0,0)(133.8,50)
\psline[linewidth=0.6,linestyle=dotted](2.5,7.5)(2.5,45)
\psline[linewidth=0.6,linestyle=dotted](65,7.5)(65,45)
\psline[linewidth=0.6,linestyle=dotted](127.5,7.5)(127.5,45)
\psline[linewidth=0.6,arrowscale=1 2,arrowsize=2 2]{|->}(2.5,30)(65,30)
\psline[linewidth=0.6,arrowscale=1 2,arrowsize=2 2]{|->}(65,30)(127.5,30)
\pscustom[linewidth=0.6]{\psbezier(127.5,30)(129.7,30)(131.12,29.56)(132.23,28.54)
\psbezier(133.33,27.53)(133.8,26.22)(133.8,24.18)
\psbezier(133.8,22.13)(133.8,20.39)(133.8,18.34)
\psbezier(133.8,16.3)(133.33,14.99)(132.23,13.97)
\psbezier(131.12,12.95)(130.18,12.51)(129.08,12.51)
\psbezier(127.97,12.51)(127.5,12.51)(127.5,12.51)
\psbezier(127.5,12.51)(127.5,12.51)(127.5,12.51)
\psbezier(127.5,12.51)(127.5,12.51)(127.5,12.51)
}
\psline[linewidth=0.6,arrowscale=1 2,arrowsize=2 2]{|->}(127.5,12.5)(65,12.5)
\psline[linewidth=0.6,arrowscale=1 2,arrowsize=2 2]{|->}(65,12.5)(2.5,12.5)
\rput[b](2.5,2.5){$t_\text{A}$}
\rput[b](65,2.5){$t_\text{B}$}
\rput[b](127.5,2.5){$t_\text{C}$}
\rput[bl](5,40){Benettin steps, store}
\rput[bl](5,35){$\bkwmat$ and $\mat{R}^\supF$}
\rput[bl](5,22.5){Iterate with $(\mat{R}^\supF)^{-1}$,}
\rput[bl](5,17.5){compute CLVs}
\rput[bl](67.5,40){Benettin steps, store $\mat{R}^\supF$}
\rput[bl](67.5,22.5){Iterate with $(\mat{R}^\supF)^{-1}$}
\rput[l](0,50){b)}
\end{pspicture}
  \end{center}
  \caption{Computation of covariant Lyapunov vectors (CLVs). a)
      Method of LU factorization (see
      Sec.~\ref{sec:lu-factorization}), and orthogonal complement
      method of Wolfe and Samelson (see
      Sec.~\ref{sec:compl-subsp}). b) Iterative method of Ginelli et
      al. (see Sec.~\ref{sec:iterations}).}
  \label{fig:methods}
\end{figure*}

Normally, one has to compute the covariant Lyapunov vectors for a
series of subsequent points of a trajectory. A practical
implementation of the algorithm in this case can be the following. We
start the procedure for Lyapunov exponents forward in time including
the iterations with $\fpropag(t_1,t_2)$ and QR factorizations, and
perform it as long as required for the orthogonal matrices
$\mat{Q}(t)$ to converge to the matrices of the backward Lyapunov
vectors $\bkwmat(t)$. Denote the end of the preliminary stage as
$t_{\text{A}}$. After this point the iterations are continued, but now
we store trajectory points of the basic system and the backward
vectors $\bkwmat(t_n)$, see the diagram in
Fig.~\ref{fig:methods}(a). The duration of this stage depends on the
number of points where we need to know the covariant vectors. At
$t_{\text{B}}$ we stop the storing of $\bkwmat(t_n)$ and, moreover,
stop the procedure for Lyapunov exponents and continue to solve only
the basic system saving the trajectory points. This stage lasts from
$t_{\text{B}}$ to $t_{\text{C}}$. Its duration must be long enough for
the subsequent backward procedure to converge. At $t_{\text{C}}$ we
start moving back along the saved trajectory performing the backward
procedure for Lyapunov exponents including iterations with the adjoint
propagator $\gpropag^{-1}$ and QR factorizations. Upon the arrival at
$t_{\text{B}}$ we have the forward Lyapunov vectors $\fwdmat(t)$. Now
we pass the interval from $t_{\text{B}}$ to $t_{\text{A}}$ given both
the backward vectors $\bkwmat(t_n)$, that were saved in the course of
the forward pass, and the forward vectors $\fwdmat(t_n)$. These
vectors can be used to compute the matrices $\mat{A}^-(t_n)$ by means
of $\mat{P}$ (Eq.~\eqref{eq:def_mat_p}), as explained before. In turn,
these matrices can be used to find the covariant vectors
$\clmatfwd(t_n)$, according Eq.~\eqref{eq:clv_defin}. Note that it is
not necessary to perform this procedure with the whole set of
vectors. To compute $j$ first covariant vectors we need $j$ first
backward vectors and $j-1$ first forward vectors. In the appendix we
provide a pseudocode implementation of the presented algorithm.

Columns of $\mat{A}^-(t_n)$ can also be considered as covariant
Lyapunov vectors written with respect to the basis $\bkwmat(t_n)$. The
covariant vectors in the form of $\mat{A}^-(t_n)$ have a mutual
orientation that is identical to $\clmatfwd(t_n)$. Therefore, if for
example the angles between covariant Lyapunov vectors are required,
they can be computed with respect to columns of $\mat{A}^-(t_n)$. This
allows us to save some machine time.

The numerical implementation of the described procedure includes well
established numerical routines. To perform the forward procedure for
Lyapunov exponents, besides of numerically solving the dynamical
equations, one also needs to compute QR decompositions. For
high-dimensional systems good results are obtained with algorithms
based on Householder transformations~\cite{CompMeth,GolubLoan}. The
backward steps may in addition require an interpolation of the stored
trajectory to find a solution of variational equations with variable
time steps.  Finally, each column of $\mat{A}^-(t_n)$ is the null
space of a corresponding rectangular submatrix of $\mat{P}$. One of
the most reliable methods of computation of the null space is based on
the SVD~\cite{GolubLoan}. The null vector is identified as a right
singular vector corresponding to the vanishing singular value.  Above
we discussed that in principle in the case of degeneracy of Lyapunov
exponents one can obtain more than one null vector for one column
$\mat{A}^-(t_n)$. But exactly identical Lyapunov exponents are
unlikely to occur in numerical computations, and, hence, multiple null
vectors can (practically) never appear.  It means that among right
singular vectors we always have a preferable candidate with the
smallest singular value.

Implementations of QR decomposition and SVD in Fortran can, for
example, be found in the well-known LAPACK library~\cite{LAPACK}. For a
C++ implementations we refer to the ALGLIB~NET
library~\cite{ALGLIB}. Also this library provides implementations for
many other platforms, such as Delphi and VBA.

\subsection{Orthogonal complement method of Wolfe and
  Samelson}\label{sec:compl-subsp}

One of two first methods for the efficient computation of covariant
Lyapunov vectors was suggested by Wolfe and
Samelson~\cite{WolfCLV}. Just as the LU method, their approach
utilizes the local property of the covariant vectors determined by
Eq.~\eqref{eq:clv_defin}. This equation can be written for the $j$the
vector as
\begin{align}
  \label{eq:clv_bkw_decomp}
  \clvecfwd_j&=\sum_{i=1}^j\bkwvec_i\alpha^-_{ij},\\
  \label{eq:clv_fwd_decomp}
  \clvecfwd_j&=\sum_{i=j}^m\fwdvec_i\alpha^+_{ij}.
\end{align}
As above, the time dependence is not explicitly
shown. Equating Eqs.~\eqref{eq:clv_fwd_decomp} and
\eqref{eq:clv_bkw_decomp} and multiplying them by $\fwdvec_k$ we can
find
\begin{equation}\label{eq:alpha_plus}
  \alpha^+_{kj}=\sum_{n=1}^j 
  \langle\fwdvec_k\bkwvec_n\rangle\alpha^-_{nj}.
\end{equation}
Now we substitute this $\alpha^+_{kj}$ in
Eq.~\eqref{eq:clv_fwd_decomp} and multiply the resulting equation by
$\bkwvec_k$. Taking into account that
$\langle\bkwvec_k\clvecfwd_j\rangle=\alpha^-_{kj}$, we obtain:
\begin{equation}\label{eq:alpha_minus}
  \alpha^-_{kj}=\sum_{n=1}^j 
  \left(
    \sum_{i=j}^m p_{ik}p_{in}
  \right)
  \alpha^-_{nj},\;k\leq j.
\end{equation}
where $p_{ik}=\langle\fwdvec_i\bkwvec_k\rangle$ are elements of the
matrix $\mat{P}$ \eqref{eq:def_mat_p}. 

In principle, this equation allows one to compute $\alpha^-_{kj}$ and
to find the covariant vectors via Eq.~\eqref{eq:clv_bkw_decomp}. But
this straightforward approach is not efficient. To compute the $j$th
covariant vector, the coefficients $\alpha^-_{kj}$ are required, where
$k=1,2,\ldots,j$. These coefficients depend on
$p_{ik}=\langle\fwdvec_i\bkwvec_k\rangle$, where
$i=j,j+1,\ldots,m$. So, we need $m-j+1$ last vectors $\fwdvec$, and
$j$ first vectors $\bkwvec$. The total number is always $m+1$.

The key idea of Wolfe and Samelson to avoid this obstacle utilizes the
orthogonality of $\mat P$~\cite{WolfCLV, WolfSamelsonConf}. One can
obtain the needed subspace spanned by the last $(m-j+1$) vectors by
taking the orthogonal complement to the subspace of the first $(j-1)$
vectors. In more detail, columns of $\mat{P}$ are orthogonal to each
other, i.e., $\sum_{i=1}^{m}p_{ik}p_{in}=\delta_{kn}$, where
$\delta_{kn}=1$ if $k=n$ and 0 otherwise. This sum can be split at
$i=j$ as follows:
\begin{equation}
  \sum_{i=j}^{m}p_{ik}p_{in}=\delta_{kn}-\sum_{i=1}^{j-1}p_{ik}p_{in}.
\end{equation}
The sum at the left hand side of this equation includes elements from
the last rows of $\mat{P}$, while the sum at the right hand side
consists of the elements of the first rows. So, the sum in parentheses
in Eq.~\eqref{eq:alpha_minus} can be substituted as:
\begin{multline}
  \alpha^-_{kj}=\sum_{n=1}^j 
  \left(
    \delta_{kn}-\sum_{i=1}^{j-1}p_{ik}p_{in}
  \right)
  \alpha^-_{nj}\\=
  \alpha^-_{kj}-
  \sum_{n=1}^j 
  \left(
    \sum_{i=1}^{j-1}p_{ik}p_{in}
  \right)
  \alpha^-_{nj}.
\end{multline}
Thus, to compute $j$ unknown coefficients $\alpha^-_{nj}$, where
$n\leq j$, we have to solve a set of $j$ linear homogeneous equations
\begin{equation}\label{eq:ws_find_clv}
  \sum_{n=1}^j 
  \left(
    \sum_{i=1}^{j-1}p_{ik}p_{in}
  \right)
  \alpha^-_{nj}=0 \;\; (j=1,2,\ldots,m,\; k\leq j).
\end{equation}
(We remind the reader that $\alpha^-_{nj}=0$ for $n>j$.) 
Equation~\eqref{eq:ws_find_clv} was suggested by Wolfe and Samelson to
compute $\mat{A}^-$. It does not depend on the last rows of $\mat{P}$,
so that one needs $j$ first backward vectors and $j-1$ first forward
vectors to compute $j$ first covariant vectors.

Later the method of Wolfe and Samelson was modified by Paz\'o et
al.~\cite{PazoSzendro08} using the standard approach of computation of
the forward and backward Lyapunov vectors, based on QR factorizations
and on the backward iterations with the transposed propagator (these
ideas were discussed in Sec.~\ref{sec:num_comp_lyap}).

Changing the order of sums in Eq.~\eqref{eq:ws_find_clv}, we can write
it in matrix form as
\begin{equation}\label{eq:ws_find_clv_mat}
  \mat{P}(1\mycolon j-1,1\mycolon j)^\supT\mat{P}(1\mycolon j-1,1\mycolon j)
  \mat{A}^-(1\mycolon j,j)=0.
\end{equation}
Compare this equation with Eq.~\eqref{eq:lu_find_clv}. We can see that
solutions of Eq.~\eqref{eq:lu_find_clv} constitute a subset of
solutions of Eq.~\eqref{eq:ws_find_clv_mat}. But because we need only
one solution at each $j$, and because our LU method finds such
solution, we can conclude that the LU method works in the same way as
the Wolfe and Samelson method, avoiding redundant matrix
multiplication.

\subsection{Backward iterations, method of Ginelli et
  al.}\label{sec:iterations}

Almost simultaneously with Wolfe and Samelson, Ginelli et
al.~\cite{GinCLV} suggested a method based on asymptotic properties of
covariant vectors~\eqref{eq:clv_main_prop}. The underlying idea of
this method was described in the end of
Sec.~\ref{sec:covar-lyap-vect}, but it cannot be directly
implemented. Assume that we have backward Lyapunov vectors at
$t_1$. Theoretically we can initialize $\vec{v}_j(t_1)$ satisfying
Eq.~\eqref{eq:asympt_exp_fbkw}, and start the backward iterations
using $\fpropag^{-1}$. But in practice, due to numerical noise all
these vectors shall belong to
$\fsplitminus_m(t_1)\setminus\fsplitminus_{m-1}(t_1)$, because this
set has the largest measure. Hence, these iterations can provide only
$\clvecfwd_m$. Due to the same reasons the forward iterations converge
to $\clvecfwd_1$. The same is also true for the adjoint propagator.

The key idea of Ginelli et al. is to perform the iterations in the
space of projections onto backward Lyapunov vectors $\bkwmat(t)$.  For
a set of vectors initialized according to
Eq.~\eqref{eq:asympt_exp_fbkw}, the matrix of projections onto
$\bkwmat(t)$ is upper triangular and the iterations converge in the
backward time.  As follows from Eq.~\eqref{eq:upper_triang_map}, the
backward iterations with $\fpropag(t_1,t_2)^{-1}$ in the space of
projections onto $\bkwmat(t)$ are equivalent to backward iterations
with the upper triangular matrix $\mat{R}^\supF(t_1,t_2)^{-1}$.  This
mapping preserves the triangular structure of the matrix of
projections, and we can perform as many backward iterations as we need
always staying within subspaces
$\fsplitminus_j(t_1)\setminus\fsplitminus_{j-1}(t_1)$. In other words,
any upper triangular matrix iterated backward in time with
$\mat{R}^\supF(t_1,t_2)^{-1}$ converges to $\mat{A}^-(t)$. Note that
since the subspaces $\fsplitminus_j(t)$ are spanned by the first $j$
backward Lyapunov vectors, we are allowed to compute only $j$ first
covariant vectors without computing the rest of them.

In a similar way we can compute the first $j$ adjoint covariant
vectors, using the forward-time asymptotic
\eqref{eq:asympt_exp_gfwd}. We start the procedure moving backward in
time with the transposed propagator and computing forward Lyapunov
vectors as described in Sec.~\ref{sec:num_comp_lyap}. The triangular
matrices $\mat{R}^\supG(t_1,t_2)$ have to be stored. Then we turn
round and start forward iterations
$\mat{R}^\supG(t_n,t_{n+1})^{-1}\mat{B}(t_n)=
\mat{B}(t_{n+1})[\matcadj(t_n,t_{n+1})]^{-1}$ that converge to
$\mat{B}^+(t)$.

A practical implementation of the method of Ginelli et al. might be
the following; see the illustration in
Fig.~\ref{fig:methods}(b). First, we perform the procedure for
Lyapunov exponents including forward iterations with
$\fpropag(t_1,t_2)$ and QR factorizations. This stage is preliminary
and it is finished at $t_{\text{A}}$ when we decide that the
orthogonal matrices $\mat{Q}(t)$ have converged to the matrices of
backward Lyapunov vectors $\bkwmat(t)$.  Starting from $t_{\text{A}}$,
we continue the procedure, but now all the matrices $\bkwmat(t_n)$ and
$\mat{R}^\supF(t_n,t_{n+1})$, see Eq.~\eqref{eq:fmap_bkwvec}, are
stored. This stage continues until $t_{\text{B}}$. The length of this
stage depends on the number of points where we later want to compute
the covariant vectors. After $t_{\text{B}}$ we still proceed with the
procedure, but store only $\mat{R}^\supF(t_n,t_{n+1})$. This stage
must be sufficiently long to provide the convergence of the subsequent
backward procedure and it finishes at $t_{\text{C}}$. At this point we
initialize a set of arbitrary vectors, for which the
property~\eqref{eq:asympt_exp_fbkw} is fulfilled. In fact we just
generate a random upper triangular matrix $\mat{A}$.  Using the stored
matrices $\mat{R}^\supF(t_n,t_{n+1})$, we perform the backward
iterations on the interval from $t_{\text{C}}$ to $t_{\text{B}}$.
\begin{equation}\label{eq:ginelli_iter}
  \mat{R}^\supF(t_n,t_{n+1})^{-1}\mat{A}(t_{n+1})=
  \mat{A}(t_n)\mat{C}(t_n,t_{n+1})^{-1},
\end{equation}
where the diagonal matrix $\mat{C}(t_n,t_{n+1})^{-1}$ contains column
norms of $\mat{A}$. If $t_{\text{C}}-t_{\text{B}}$ is sufficiently
large, $\mat{A}(t_n)$ converges to $\mat{A}^-(t_n)$. Now we pass the
stage from $t_{\text{B}}$ to $t_{\text{A}}$ computing the covariant
Lyapunov vectors via Eq.~\eqref{eq:clv_defin} and using them as we
need. Note, that this procedure allows one to compute not only the whole
set of $m$ covariant vectors, but also as many of them as we want.

As we already mentioned above, the columns of $\mat{A}^-(t_n)$ can
also be considered as covariant Lyapunov vectors, so that in some
cases it is enough to consider these vectors without computation of
$\clmatfwd(t_n)$. In this case the matrices $\bkwmat(t_n)$ do not have
to be stored.

The algorithm of backward iterations can suffer from ill-conditioned
$\mat{R}^\supF$, which manifests itself if one computes many (i.e., not
just a few first) covariant Lyapunov vectors for a system with strong
contraction. Typically, high-dimensional chaotic dissipative systems
have several positive Lyapunov exponents of moderate magnitude while
negative exponents can have large absolute values.  Because logarithms
of diagonal elements of $\mat{R}^\supF$ are proportional to local
Lyapunov exponents, they can be sufficiently small. So, if a lot of
covariant vectors corresponding to negative Lyapunov exponents are
computed, the diagonal elements of $\mat{R}^\supF$ can become small,
and the whole matrix $\mat{R}^\supF$, whose determinant is the product
of its diagonal elements, can potentially be ill-conditioned. In turn
this can influence the accuracy of computations.

To avoid or at least minimize this problem one should first try to
decrease the interval between QR orthogonalizations. Another, also
almost obvious recommendation is not to employ
Eq.~\eqref{eq:ginelli_iter} as it is, but compute iterations
implicitly. Note that the implicit method is preferable regardless of
the presence of ill-conditioned $\mat{R}^\supF$. Namely, nonzero
elements of the $i$th column of $\mat{A}(t_n)$ can be computed as a
solution of equation
\begin{equation}\label{eq:ginelli_iter_impl}
  \mat{R}^\supF(1\mycolon i,1\mycolon i)\mat{A}_n(1\mycolon i,i)
  =\mat{A}_{n+1}(1\mycolon i,i),
\end{equation}
where $\mat{R}^\supF(1\mycolon i,1\mycolon i)$ is a top left submatrix
of $\mat{R}^\supF$ and $\mat{A}_n(1\mycolon i,i)$ top fragment of the
$i$th column of $\mat{A}(t_n)$. Computed in this way
$\mat{A}_n(\mycolon,i)$ then has to be normalized. We see that the
$i$th column of $\mat{A}(t_n)$ is influenced only by the submatrix
$\mat{R}^\supF(1\mycolon i,1\mycolon i)$ that remains well-conditioned
until $i$ is sufficiently small. It means that even if $\mat{R}^\supF$
has some small diagonal elements, errors that they can produce are not
spread along the whole spectrum, but influence only minor covariant
vectors from its right part.

When the trajectory passes close to tangencies of invariant manifolds
of an attractor, $\mat{A}(t_n)$ becomes ill-conditioned, i.e., small
values can appear on its diagonal. Because $\mat{A}(t_n)$ is used to
compute $\mat{A}(t_{n-1})$, small values can accumulate and vanish due
to the numerical underflow. Then the zeros will be preserved in the
course of iterations even if the trajectory goes far from the tangency
points. This false indication of an exact tangency can be cured by
adding a small amount of noise to the diagonal elements.

\subsection{Comparison of the methods}

Computation of covariant vectors requires saving of intermediate
matrices. We estimate the amount of the required memory for the
``worst'' case when the whole set of $m$ covariant vectors is
computed. Let $\stpcomp$ be the number of trajectory points where we
are going to compute covariant vectors, i.e., the number of steps in
the stage AB in Fig.~\ref{fig:methods}. It is reasonable to assume
that this value depends on $m$, $\stpcomp=\stpcomp(m)$, where $m$ is
the dimension of the phase space. Denote the number of steps in the
transient stage BC by $\stptran$. The convergence of columns of
matrices to their asymptotic form during the transient stage is
exponential with rates equal to differences between corresponding
Lyapunov exponents~\cite{WolfCLV}. For extensive chaotic systems these
differences are proportional to $1/m$; thus, the convergence time is
proportional to $m$. Altogether, the length of the transient stage can
be estimated as $\stptran=\stptrancoef m$, where $\stptrancoef$ is an
empirical constant, which depends on the particular system under
consideration.

For the LU method, Sec.~\ref{sec:lu-factorization}, and for the method
of orthogonal complement, Sec.~\ref{sec:compl-subsp}, the estimates
are identical. Namely, we need $\stpcomp$ matrices $\bkwmat$, each of
the size $m^2$, and $\stpcomp+\stptran$ trajectory vectors of the size
$m$, see Fig.~\ref{fig:methods}(a). Hence, the total amount of memory
(in bytes) is
$B_{\text{LU}}=(m^2(\stpcomp(m)+\stptrancoef)+m\stpcomp(m))b$, where
$b$ is the number of bytes required to store one real number. For
large $m$ we have
\begin{equation}\label{eq:lu_ram}
  B_{\text{LU}}\approx m^2\stpcomp(m)b.
\end{equation}
For example if the dimension is $m=100$ and we want to compute
$\stpcomp=1000$ covariant vectors using double precision numbers,
i.e., $b=8$, we need $B_{\text{LU}}\approx 76\text{ megabytes}$.

For the method of backward iterations, Sec.~\ref{sec:iterations}, we
need to save $\stpcomp+\stptran$ triangular matrices $\mat{R}^\supF$,
each of the size $(m^2+m)/2$, and $\stpcomp$ matrices $\bkwmat$ of the
size $m^2$, see Fig.~\ref{fig:methods}(b). The total amount of memory
can be estimated as $B_{\text{BI}}=(m^2(3\stpcomp(m)+\stptrancoef
m)+m(\stpcomp(m)+\stptrancoef m))b/2$. Keeping only the leading terms
for large $m$ we obtain:
\begin{equation}\label{eq:gin_ram}
  B_{\text{BI}}\approx m^2(3\stpcomp(m)+\stptrancoef m)b/2.
\end{equation}
For the same numerical values as in the example for LU method and at
$\stptrancoef=1$ we obtain, though higher, but close estimate:
$B_{\text{BI}}\approx 118\text{ megabytes}$. Note however, that the
amount of memory for the transient stage grows with $m$ as
$\stptrancoef m^3b/2$ for the backward iterations method, while for
two other methods it grows as $\stptrancoef m^2 b$. Hence, the
efficient application of the backward iterations requires closer
attention to the minimization of the transient stage length,
otherwise, one can easily exhaust the available memory.

In principle, all methods may suffer from a shortage of memory.  One
possible way to handle this problem is to save intermediate data to
binary files. The disadvantage of this approach is deceleration of
computations due to the slowness of file operations. Alternatively,
see Ref.~\cite{GinCLV}, instead of keeping all necessary matrices
moving forward in time, one can periodically (and sufficiently seldom
to fit in the available memory) save snapshots of the procedure for
Lyapunov exponents (i.e., the trajectory points of the basic system
together with corresponding matrices $\bkwmat$). Then, moving
backward, one periodically uses these snapshots to recompute forward
steps and obtain missing data. Of course, this approach also slows
down the computations, now due to the recomputations. To choose the
preferable way one has to compare the average time for writing to file
and subsequent reading of one matrix with the time needed to recompute
it. The result of comparison depends on the particular computer
system. Note also that using the method of backward iterations one can
reduce the memory consumption if only the angles between covariant
vectors are needed. As we already mentioned in
Sec.~\ref{sec:lu-factorization}, the triangle matrices $\mat{A}^-$ are
suitable for finding the angles, and hence, in this case one does not
need to save matrices $\bkwmat$.

Let us estimate the computation speed of the methods presented (the
straightforward intersection of the Oseledec subspaces is not taken
into account). If all the methods have enough memory to avoid either
using files or performing recomputing, the backward iterations are the
fastest. Local methods of LU factorization and orthogonal complement
loose the race on the backward stage B-A, see
Fig.~\ref{fig:methods}. Each iteration is simultaneously a time step
and also a computation of the covariant vectors. The time steps for
local methods are performed via the procedure for Lyapunov exponents
and also some time is required to compute the covariant vectors.

\section{Examples}\label{sec:examp}  

\subsection{System with constant Jacobian matrix} 

Consider a system with a constant Jacobian matrix
\begin{equation}
  \mat{J}=
  \begin{pmatrix}
    1 & -2 & 0\\
    0 & -1 & 0\\
    0 &  2 & -3
  \end{pmatrix}.
\end{equation}
Since $\mat{J}$ is time-independent and has real eigenvalues, the
Lyapunov exponents for this system simply coincide with the magnitude
of its eigenvalues, $\lambda_{1,2,3}=1,-1,-3$. The corresponding
eigenvectors are simultaneously the covariant Lyapunov vectors, and
the eigenvectors of $(-\mat{J}^\supT)$ are the adjoint covariant
vectors:
\begin{equation}\label{eq:examp_eigen}
  \clmatfwd=
  \begin{pmatrix}
    1 & \sqrt{1/3} & 0\\
    0 & \sqrt{1/3} & 0\\
    0 & \sqrt{1/3} & 1
  \end{pmatrix},
  \clmatadj=
  \begin{pmatrix}
    \sqrt{1/2}  & 0 & 0 \\
    -\sqrt{1/2} & 1 & -\sqrt{1/2} \\
    0 & 0 & \sqrt{1/2}
  \end{pmatrix}.
\end{equation}
$\signat=\clmatadj^\supT\clmatfwd=
\diagmat[\sqrt{1/2},\sqrt{1/3},\sqrt{1/2}]$. The propagator reads:
\begin{equation}
  \fpropag(t_1,t_2)=
  \clmatfwd\mat{L}\clmatfwd^{-1}=
  \begin{pmatrix}
    \myexp{\tau} & \myexp{-\tau}(1-\myexp{2\tau}) & 0 \\
    0 & \myexp{-\tau} & 0 \\
    0 & \myexp{-3\tau}(\myexp{2\tau}-1) & \myexp{-3\tau}
  \end{pmatrix},
\end{equation}
where $\tau=t_2-t_1$, and
$\mat{L}=\diagmat[\myexp{\lambda_1\tau},\myexp{\lambda_2\tau},\myexp{\lambda_3\tau}]$.
Forward and backward Lyapunov vectors can be computed as eigenvectors
of far-future and far-past operators, respectively, directly from
Eqs.~\eqref{eq:far_fwd_oper} and \eqref{eq:far_pst_oper} (finding the
limits one has to keep constant norms of vectors):
\begin{equation}
  \bkwmat=
  \begin{pmatrix}
    1 & 0 & 0 \\ 
    0 & \sqrt{1/2} & -\sqrt{1/2}\\
    0 & \sqrt{1/2} & \sqrt{1/2}
  \end{pmatrix},
  \fwdmat=
  \begin{pmatrix}
    \sqrt{1/2} & \sqrt{1/2} & 0 \\
    -\sqrt{1/2} & \sqrt{1/2} & 0 \\
    0 & 0 & 1
  \end{pmatrix}.
\end{equation}
Note, that in accordance with Eq.~\eqref{eq:clv_defin}, the first
backward vector $\{1,0,0\}$ and the last forward Lyapunov vector
$\{0,0,1\}$, coincide with the first and the last covariant vectors,
i.e., with eigenvectors of $\mat{J}$.  One can also check that the
logarithms of eigenvalues of the limit operators, i.e., the Lyapunov
exponents, indeed coincides with the magnitude of the
eigenvalues of $\mat{J}$.
The matrix $\mat{P}$, as defined by Eq.~\eqref{eq:def_mat_p}, reads:
\begin{equation}
  \mat{P}=
  \begin{pmatrix}
    \sqrt{1/2} & -1/2 & 1/2\\
    \sqrt{1/2} & 1/2 & -1/2\\
    0 & \sqrt{1/2} & \sqrt{1/2}
  \end{pmatrix}.
\end{equation}  

To compute covariant vectors via the LU method, we have to find the
matrix $\mat{A}^-$. As follows from Eq.~\eqref{eq:lu_find_clv}, the
first column of this matrix is always $\{1,0,0\}$ while for the other
elements we have 
$a^-_{12}/\sqrt{2}-a^-_{22}/2=0$,
$a^-_{13}/\sqrt{2}-a^-_{23}/2+a^-_{33}/2=0$, and
$a^-_{13}/\sqrt{2}+a^-_{23}/2-a^-_{33}/2=0$. For the matrix
$\mat{B}^+$, needed to compute the adjoint covariant vectors, we
construct equations according to Eq.~\eqref{eq:lu_find_clv_adj} using
$\mat{P}^\supT$: $b_{12}/\sqrt{2}+b_{22}/\sqrt{2}=0$,
$b_{13}/\sqrt{2}+b_{23}/\sqrt{2}=0$,
$-b_{13}/2+b_{23}/2+b_{33}/\sqrt{2}=0$. Both of these equation sets
have to be solved with the additional requirement of unit column
norms:
\begin{equation}
  \label{eq:exampl_aminus}
  \mat{A}^-=
  \begin{pmatrix}
    1 & \sqrt{1/3} & 0 \\
    0 & \sqrt{2/3} & \sqrt{1/2} \\
    0 & 0 & \sqrt{1/2}
  \end{pmatrix},
  \mat{B}^+=
  \begin{pmatrix}
    1 & -\sqrt{1/2} & 1/2 \\
    0 & \sqrt{1/2} & -1/2 \\
    0 & 0 & \sqrt{1/2}
  \end{pmatrix}.
\end{equation}
One can check that Eqs.~\eqref{eq:clv_defin} and
\eqref{eq:clv_defin_adj} are fulfilled, i.e.,
$\clmatfwd=\bkwmat\mat{A}^-$ and $\clmatadj=\fwdmat\mat{B}^+$. 

The method of Wolfe and Samelson does essentially the same
job. Computing $\mat{A}^-$ we have to multiply submatrices of
$\mat{P}$ by the transposed submatrices and construct equations; see
Eq.~\eqref{eq:ws_find_clv_mat}. Similarly one can get $\mat{B}^+$ and
verify that the results coincide with Eq.~\eqref{eq:exampl_aminus}.

For the method of Ginelli et al. we find
$\mat{R}^\supF(t_1,t_2)=[\bkwmat]^\supT\fpropag(t_1,t_2)\bkwmat$; see
Eq.~\eqref{eq:fmap_bkwvec}. Since the iterations
\eqref{eq:ginelli_iter} converge in backward time, consider
$\mat{R}^\supF(t_1,t_2)^{-1}$:
\begin{equation}
  \mat{R}^\supF(t_1,t_2)^{-1}=
  \begin{pmatrix}
    \myexp{-\tau} & (\myexp{\tau}-\myexp{-\tau})/\sqrt{2} & 
    (\myexp{-\tau}-\myexp{\tau})/\sqrt{2} \\
    0 & \myexp{\tau} & \myexp{3\tau}-\myexp{\tau} \\
    0 & 0 & \myexp{3\tau}
  \end{pmatrix}.
\end{equation}
As follows from Eq.~\eqref{eq:ginelli_iter}, at $\tau\to\infty$ the
column norms of $\mat{R}^\supF(t_1,t_2)^{-1}$ have to grow as
$\myexp{-\lambda_i\tau}$. Indeed, it can be checked that the column
norms of this matrix are asymptotically dominated by the terms
$\myexp{-\tau}$, $\myexp{\tau}$, and $\myexp{3\tau}$, respectively. If
we normalize columns to the unit, the elements of this matrix converge
to $\mat{A}^-$, see Eq.~\eqref{eq:exampl_aminus}, i.e., we again
obtain the covariant vectors.

\subsection{Generalized H\'enon map} \label{sec:Henon}

As second example we consider a generalized  three-dimensional H\'enon map
\cite{BK90}

\begin{equation}\label{eq:BKMap}
  \begin{aligned}
    x_1^{n+1 } & =  a - \left[  x_2^n \right]^2 - b x_3^n \\
    x_2^{n+1}  & =  x_1^n    \\
    x_3^{n+1}  & =  x_2^n.
  \end{aligned}
\end{equation}

For $a=1.76$ and $b=0.1$ this system generates a hyperchaotic
attractor with Lyapunov exponents $\lambda_1 = 0.225$, $\lambda_2 =
0.188$, and $\lambda_3 = -2.716 $.

\begin{figure*}
  \begin{center}
   \includegraphics{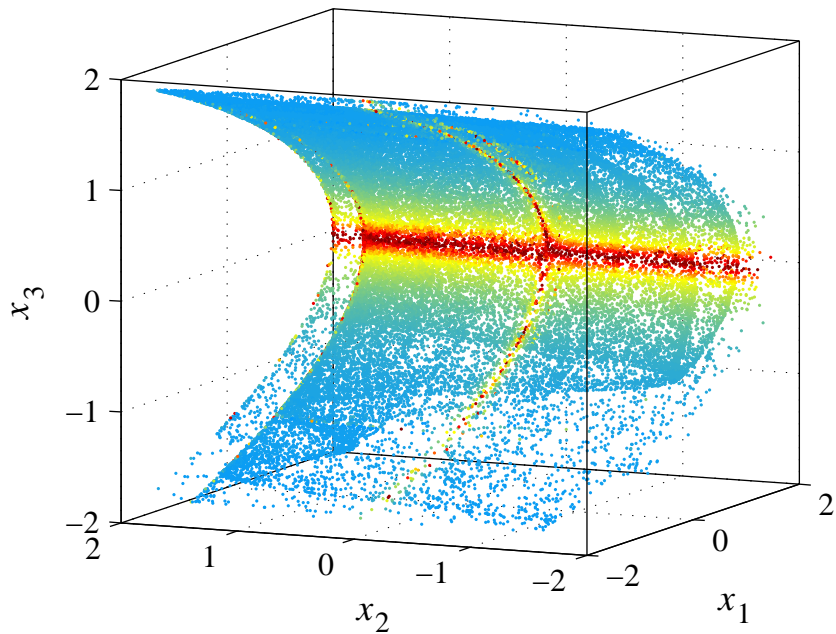}
  \end{center}
  \caption{Attractor of the generalized H\'enon map
    Eq.~\eqref{eq:BKMap}. Dark (red) colors indicate closeness to
    homoclinic tangencies. (Color figure online)}
  \label{fig:Henon}
\end{figure*}

Figure~\ref{fig:Henon} shows the chaotic attractor, where the color of
the points corresponds to $\det[\mat P(1\mycolon 2,1\mycolon 2)]$ (see
Sec.~\ref{sec:lu-factorization}).  Dark (red) colors indicate
locations of the attractor where (almost) tangent CLVs occur and the
submatrix $\mat P(1\mycolon j,1\mycolon j)$ with $j=2$ is (almost)
singular.

\section{Conclusion}\label{sec:conclusion}

We presented an extensive description of modern achievements of
Lyapunov analysis. The Lyapunov exponents, the forward and backward
Lyapunov vectors as well as covariant Lyapunov vectors were discussed
in detail. 

The systematic approach allowed us to reveal a symmetry in the
structure of the tangent space and to introduce the concept of adjoint
covariant vectors. There are tangent linear propagators that can be
characterized by left and right singular vectors. When the propagators
are considered on asymptotically growing time intervals these singular
vectors converge to backward and forward Lyapunov vectors. One can
also define adjoint propagators that are associated with the same
singular vectors, but have reciprocal singular values.  The backward
and forward Lyapunov vectors can be used as frameworks for two sets of
Oseledec subspaces and for two adjoint Oseledec subspaces that are
orthogonal to the Oseledec subspaces.  The main feature of these
subspaces is the covariance with the tangent dynamics: the propagator
maps each Oseledec subspace onto the corresponding Oseledec subspace
associated with the image point of the trajectory, and the adjoint
propagator does the same with the adjoint subspaces. Within these
subspaces one can find vectors with the same property of
covariance. There are covariant Lyapunov vectors whose exponential
growth under the action of the propagators is characterized by
Lyapunov exponents, and there are also adjoint covariant Lyapunov
vectors that grow under the action of adjoint propagators with
Lyapunov exponents of opposite signs.

The adjoint covariant vectors are not independent characteristic
vectors, because in principle one can always compute them using the
original covariant Lyapunov vectors.  Important are the
norm-independent angles between corresponding covariant and adjoint
vectors. They provide a compact representation of the information
provided by covariant vectors. In particular, homoclinic tangencies
between stable and unstable manifolds (characteristic for
non-hyperbolic chaos) are indicated by orthogonality of corresponding
original and adjoint vectors.

An important result of our detailed analysis is an efficient method
for computing covariant Lyapunov vectors. The basic idea of the method
is an optimized LU decomposition of the matrix $\mat{P}$ consisting of
scalar products of forward and backward Lyapunov vectors. Our approach
is very close to the method by Wolfe and Samelson~\cite{WolfCLV}, but
its advantages are a more transparent explanation, and the explicit
formulation of the matrix $\mat{P}$ which is interesting by
itself. Moreover our approach is slightly more efficient because we
avoid some redundant computations.

Using the matrix $\mat{P}$, we present a method for detecting
non-hyperbolicity of chaotic dynamics without explicit computation of
the covariant vectors. In brief, the violation indicator is the
singularity of a $j\times j$ submatrix of $\mat{P}$, where $j$ is the
number of positive Lyapunov exponents. The chaotic dynamics is
non-hyperbolic if moving along a trajectory we encounter nearly
singular submatrices.

In presence of degenerate Lyapunov exponents all types of Lyapunov
vector are not unique. We provide an analysis of this case. As for
the forward and backward Lyapunov vectors, the standard algorithms can
be used without modifications. Selection of an orthogonal initial
matrix eliminates the ambiguity. Starting from different seed
matrices, we can obtain different sets of vectors, but any one of them is
appropriate. Moreover, in practical computations the degeneracy of the
Lyapunov exponents manifests itself very weakly, especially for
systems of high dimension. Typically, due to numerical errors all
computed exponents are distinct, and one cannot identify degenerate
exponents just by examining the computed spectrum. The same is
true for the covariant vectors. Theoretically the degeneracy of the
Lyapunov exponents can result in multiple sets of covariant vectors,
but in practice the computations can be organized in a such way that
one always obtains a unique appropriate solution regardless of the
degeneracy.

\begin{acknowledgments}
  The research leading to the results has received funding
  from the European Community's Seventh Framework Programme
  FP7/2007–2013 under grant agreement No. HEALTH-F2-2009-241526,
  EUTrigTreat.  P.~V.~K. acknowledges support from RFBR-DFG under
  Grant No. 08-02-91963.
\end{acknowledgments}

\appendix*

\section{Pseudocode for the LU method}

{\bf Input:} \verb'nclv', number of computed covariant Lyapunov
vectors; \verb'nstore', number of trajectory points where the
covariant vectors are computed; \verb'm', dimension of the tangent
space; \verb'dt', time interval between orthogonalizations
(normally, a multiple of time discretization step); \verb'nspend_att',
\verb'nspend_fwd', \verb'nspend_bkw', steps to converge to the
attractor, forward and backward vectors, respectively.

{\bf Subroutines:} \verb'solve_bas()', solving of the basic system;
\verb'solve_lin_fwd()', \newline \verb'solve_lin_trp()', action of forward
and transposed propagators, respectively (see
Sec.~\ref{sec:num_comp_lyap}); \verb'null_vect()', computing a null
vector (in the case of multiple solutions, an arbitrary null-vector
can be taken); \verb'orthog()', QR-orthogonalization (matrix
$\mat{R}$ is abandoned); \verb'transpose()', transpose of a matrix;
\verb'random()', generate random matrix or vector; \verb'A.B',
multiplication of matrices \verb'A' and \verb'B'.

{\bf Result:} \verb'Gamma', array of \verb'nstore' matrices \verb'm' by
\verb'nclv', whose columns are the covariant Lyapunov vectors.

\begin{verbatim}
BEGIN clv_lu
  // *** ARRIVE AT THE ATTRACTOR ***
  CREATE u[1:m]=random(1,m)
  u=solve_bas(u,dt*nspend_att)
  // *** PRELIMINARY STAGE ***
  CREATE Q[1:m][1:nclv]=random(1,m,1,nclv)
  Q=orthog(Q)
  FOR i=1 TO nspend_fwd
    Q=solve_lin_fwd(Q,u,dt)
    Q=orthog(Q)
    u=solve_bas(u,dt)
  NEXT i
  // *** STAGE A-B ***
  CREATE PhiMns[1:nstore][1:m][1:nclv]
  CREATE traj[1:nstore+nspend_bkw][1:m]
  FOR i=1 TO nstore
    Q=solve_lin_fwd(Q,u,dt)
    Q=orthog(Q)
    u=solve_bas(u,dt)
    traj[i]=u
    PhiMns[i]=Q
  NEXT i
  // *** STAGE B-C ***
  FOR i=1 TO nspend_bkw
    u=solve_bas(u,dt)
    traj[nstore+i]=u
  NEXT i
  // *** STAGE C-B *** 
  // Now we use one column less
  RECREATE Q[1:m][1:nclv-1]=random(1,m,1,nclv-1)
  Q=orthog(Q)
  // We leave this cycle at 
  // the (nstore+1)th trajectory point!
  FOR i=nspend_bkw TO 2 STEP -1
    u=traj[nstore+i]
    Q=solve_lin_trp(Q,u,dt)
    Q=orthog(Q)
  NEXT i
  // *** STAGE B-A ***
  CREATE P[1:nclv-1][1:nclv]
  CREATE Gamma[1:nstore][1:m][1:nclv]
  CREATE a[1:nclv]
  // We come into this cycle being at 
  // the (nstore+1)th point 
  // and take traj[i+1], but not traj[i].
  FOR i=nstore TO 1 STEP -1
    u=traj[i+1]
    Q=solve_lin_trp(Q,u,dt)
    Q=orthog(Q)
    P=transpose(Q).PhiMns[i]
    Gamma[i][1:m][1]=PhiMns[i][1:m][1]
    FOR j=2 TO nclv
      a[1:j]=null_vect(P[1:j-1][1:j])
      Gamma[i][1:m][j]=PhiMns[i][1:m][1:j].a[1:j]
    NEXT j
  NEXT i
END 
\end{verbatim}

\bibliography{lv}

\end{document}